\documentclass{aa}
\usepackage{upgreek}
\usepackage{graphicx}
\usepackage[varg]{txfonts}
\usepackage{natbib}
\usepackage{float}
\usepackage{soul}
\newlength{\linwx}
\usepackage{color}

\begin{document}

\title{The origin of the high heavy-element content of warm Jupiters}
\author{
Bertram Bitsch
}
\offprints{B. Bitsch,\\ \email{bbitsch@ucc.ie}}
\institute{
Department of Physics, University College Cork, Cork, T12 R229, Ireland
}
\abstract{Interior models of planets have revealed that warm Jupiters should harbour high heavy-element masses, with Jupiter mass planets potentially consisting of up to 30-40\% of heavy elements (heavier than hydrogen and helium). However, it is unclear whether the heavy-element mass of giant planets decreases with decreasing stellar mass. Within the pebble drift and evaporation scenario, inward-drifting pebbles can evaporate, enriching the inner disc with vapour to super-stellar values, which can then be accreted by growing giant planets. Due to a sequential evaporation sequence -- where H$_2$O evaporates in the inner disc, while CO$_2$, CH$_4$, and CO evaporate in the outer disc -- the vapour enrichment of the disc decreases with increasing orbital distance. We employed the \textsc{chemcomp} code to model the migration and growth of planets via pebble and gas accretion in discs around stars of different masses and discs of different sizes. Our simulations show that: (i) pebble drift and evaporation can explain the wide range of derived heavy-element content in giant planets; (ii) giant planets forming around stars with larger disc sizes can accrete larger quantities of heavy elements, because these discs can maintain a high pebble flux for a longer time, allowing the inner discs to be enriched for longer and resulting in planets with higher heavy-element content; (iii) planets forming or migrating interior to the water ice line can accrete larger amounts of heavy elements via the accretion of vapour-enriched gas compared to those forming in the outer disc regions; (iv) metal-rich stars can enable the formation of planets more enriched in heavy elements if they form in the inner disc regions, whereas planets forming in the outer disc regions -- regardless of disc metallicity -- are much less enriched than their inner-disc  counterparts that accrete vapour-enriched gas; and (v) planets forming in discs with the same size and total mass but around higher-mass stars are only marginally more enriched than those forming around lower-mass stars. Together, our study suggests that the trend of increasing total heavy-element mass with increasing planetary mass arises from the accretion of vapour-enriched gas in discs of different sizes, combined with the different origins of warm giant planets from the inner and outer disc.
}
\authorrunning{Bitsch}\maketitle

\section{Introduction}
\label{sec:intro}

The first detection of a planet around a main-sequence star revealed a new class of planets: hot Jupiters \citep{1995Natur.378..355M}. Since this detection, many more hot and warm Jupiters have been found, and we are now in an era in which we can characterize these planets in greater detail. While we can determine the atmosphere's composition through ground-based (e.g. \citealt{2021AJ....162...73P, 2024AJ....168..133P}) and space-based observations (e.g. \citealt{2022arXiv221110489A, 2023arXiv230507753A, 2025NatAs...9..845E}), we can also determine the total heavy-element budget of these planets using interior models (e.g. \citealt{2016ApJ...831...64T, 2023MNRAS.tmp.1812B, 2025A&A...693L...4M, 2025ApJ...994...43C}).

Surprisingly, these warm Jupiters can exhibit very high heavy-element content, with average values of 50-60 Earth masses for Jupiter-mass planets (e.g. \citealt{2016ApJ...831...64T, 2023MNRAS.tmp.1812B, 2025A&A...693L...4M, 2025ApJ...994...43C}). These values may even exceed the highest estimates of our own Jupiter (e.g. \citealt{2017GeoRL..44.4649W, 2018A&A...610L..14V, 2022Icar..37814937H, 2023A&A...672A..33H}). These studies also indicate that the average heavy-element content of planets increases with planetary mass. Furthermore, \citet{2026A&A...706A..50P} show that this trend holds under different assumptions for the interior structure (e.g. two-layer core and envelope, fully convective, or dilute core), demonstrating the robustness of this trend.

An additional trend has recently emerged: \citet{2025A&A...693L...4M} show that warm Jupiters around M dwarfs may be less enriched in heavy elements than their equal-mass counterparts around FGK-type stars. Given that a certain amount of heavy elements is necessary for planet formation, it is surprising that the overall enrichment of these planets around M dwarfs is lower. However, this trend was not confirmed in the study by \citet{2025ApJ...994...43C}, raising the question of whether the heavy-element mass of giant planets is truly a function of stellar mass.

As it is well known that stars enriched in heavy elements host Jupiters more frequently (e.g. \citealt{2004A&A...415.1153S, 2005ApJ...622.1102F, J2010}) due to the more efficient formation of planetesimals and planets, one could speculate that warm Jupiters forming around metal-rich stars should be more metal-rich as well (as they could accrete more planetesimals and other solids). However, the observations of \citet{2019AJ....158..239T} do not support this hypothesis. They instead show no real correlation between the heavy-element mass fraction of warm Jupiters forming around metal-rich stars compared to those forming around metal-poor stars.

The origin of this high heavy-element content is not entirely known despite the existence of different models. One set of models focuses on the accretion of several planetesimals during migration (e.g. \citealt{2020A&A...633A..33S, 2022A&A...665L...5K}), while other models focus on giant impacts between planets (e.g. \citealt{2020MNRAS.498..680G}). Alternatively, the planet could also become enriched through the accretion of gas enriched with volatile vapour (e.g. \citealt{2017MNRAS.469.3994B, 2021A&A...654A..71S, 2022A&A...665A.138B}) or small dust grains accreted with the gas (e.g. \citealt{2023arXiv230601653M, 2023A&A...679A..11B}).

This alternative scenario explaining the formation of these heavily enriched giant planets relates back to disc evolution. As millimetre--centimetre-sized particles -- so called pebbles -- form in the disc, they begin to drift inwards efficiently (e.g. \citealt{1977MNRAS.180...57W, 2008A&A...480..859B}). As these pebbles cross evaporation fronts (e.g. \citealt{2011ApJ...743L..16O}), they potentially release their volatiles into the gas phase (e.g. \citealt{2015ApJ...815..109P, 2017MNRAS.469.3994B, 2021A&A...654A..71S, 2023arXiv230815128M}). As the gas moves much more slowly than the pebbles, high levels of vapour enrichment (e.g. H$_2$O, CO$_2$, and CO) can be achieved (e.g. \citealt{2021A&A...654A..71S, 2021ApJ...921...84K, 2023arXiv230815128M}). Observations of protoplanetary discs confirm this high vapour enrichment in inner disc regions (e.g. \citealt{2023arXiv230703846B, 2023Natur.620..516P, 2024ApJ...977..173C}).

Observations of discs around M dwarfs have shown that the inner regions of these discs are enriched with carbon-bearing species, in contrast to solar-type stars, where water dominates (e.g. ~\citealt{2013ApJ...779..178P, 2023arXiv230703846B, 2024ApJ...977..173C}). This trend can be likewise explained by pebble drift and evaporation \citep{2023arXiv230815128M}: as the discs around M dwarfs are colder, the evaporation fronts of the major volatile carriers (e.g. H$_2$O, CO$_2$, CH$_4$, and CO) are located closer to their central star. Consequently, the released volatiles take less time to accrete onto the central star due to the shorter viscous timescale. As the water ice line is very close to the central star for M dwarfs, water vapour is significantly depleted within 1-2 Myr and replaced by carbon-rich vapour moving inwards from the outer disc. In contrast, this process can take even longer than 10 Myr for solar-type stars. This implies that the total vapour fraction in the inner disc regions may be higher for solar-type stars compared to M dwarfs\footnote{The study by \citet{2023arXiv230815128M} assumes that the disc size increases with stellar mass, allowing discs around higher-mass stars to maintain their pebble flux for a longer time.}.

Planets forming in the inner disc regions around solar-type stars can thus accrete large amounts of vapour, explaining their high heavy-element enrichment \citep{2017MNRAS.469.3994B, 2021A&A...654A..71S, 2022A&A...665A.138B, 2026PASJ...78..493O, 2026ApJ..1007...19B}. While the trends -- that planets accreting interior to the water ice line harbour higher heavy-element content (e.g. \citealt{2021A&A...654A..71S, 2026PASJ...78..493O}) and that planetesimal formation and accretion results in lower heavy-element content (e.g. \citealt{2023A&A...679L...7D, 2026ApJ..1007...19B}) -- seem well established, it is unclear how they scale with disc properties. In particular, it is unclear what exact role the disc's size, mass, viscosity, metallicity, and stellar mass play.

In this work, we employed the disc evolution and planet formation code \textsc{chemcomp} \citep{2021A&A...654A..71S} to study the evolution of the vapour content in protoplanetary discs and understand how this evolution affects the total heavy-element content of planets forming around stars of different masses and disc sizes. Our work is structured as follows. In Sect.~\ref{sec:methods} we discuss our methodology. In Sect.~\ref{sec:disc} we show how pebble drift and evaporation enriches inner discs around stars of different masses and sizes. In Sect.~\ref{sec:composition} we discuss the heavy-element content of individual planets and present the results for a general population in Sect.~\ref{sec:population}. We discuss the shortcomings and implications of our work in Sect.~\ref{sec:discussion} and draw our conclusions in Sect.~\ref{sec:conclusion}.

\section{Methods}
\label{sec:methods}

We used the disc evolution and planet formation code \textsc{chemcomp} \citep{2021A&A...654A..71S} to study how inward-drifting pebbles evaporate and enrich the disc with vapour and how much of the vapour is accreted onto forming planets. The code contains a viscous disc evolution model combined with pebble growth and drift, following the approach outlined in \citet{2012A&A...539A.148B}. We also trace different chemical elements within the disc and inside forming planets (see Table~\ref{tab:solar_values}). We used a solar composition \citep{2009ARA&A..47..481A} for all our simulations but varied the total metallicity across different simulations. The chemical elements were distributed into different solids (see Table~\ref{tab:comp}). We note that the amount of water is strongly affected by the amount of CO, CO$_2$, and rocky materials, as only leftover oxygen forms water. In our model we used 60\% of carbon in refractory carbon grains.

For simplicity, our model does not take the temperature evolution of the disc into account. The temperature structure of the disc is calculated at the beginning of the simulations as the balance between viscous and stellar heating and radiative cooling, and it remains constant throughout the simulation. The position of the evaporation fronts is thus determined by the stellar luminosity, as well as by the disc mass and size, which govern viscous heating. However, we expect that this does not affect the general conclusions of our study, as vapour-enriched gas would still exist interior to the inward-moving evaporation fronts. Therefore, we place our results in relation to evaporation fronts rather than orbital distances.

The planets in our models grow first via pebble accretion \citep{Johansen2017}, which stops at the pebble isolation mass when the planet opens a partial gap \citep{2014A&A...572A..35L, 2018arXiv180102341B}. This is then followed by a slow gas contraction phase that leads into runaway gas accretion \citep{2021MNRAS.501.2017N}. The contraction phase of the envelope strongly depends not only on the planetary core mass but also on the opacity within the envelope, which is regulated by the dust grains and their opacity (e.g. \citealt{2000ApJ...537.1013I, 2021A&A...647A..96B}). Depending on the assumed envelope opacity, relatively low-mass cores may transition into runaway accretion. Here, we used an envelope opacity of 0.05cm$^2$/g as in our previous simulations (e.g. \citealt{2021A&A...654A..71S}), which allows cores of $\approx$3 Earth masses to already transition into runaway accretion\footnote{Even if only cores with higher masses could transition into runaway accretion, this would not affect our results, as it would merely reduce the number of giant planets in our simulations without influencing their heavy-element content.} (see also \citealt{2026arXiv260909382P}). During their growth planets can migrate via type-I \citep{2011MNRAS.410..293P} and type-II migration \citep{2013arXiv1312.4293B}. We additionally account for heating \citep{2015Natur.520...63B} and dynamical torque effects \citep{2014MNRAS.444.2031P}, which can enable the outward migration of planets even at low viscosities (e.g. \citealt{2020arXiv200400874B}). The model is explained in detail in \citet{2021A&A...654A..71S}.

\begin{table}
 \centering
 \caption{Elemental number ratios used in our model, corresponding to the abundance of element X compared to hydrogen in the solar photosphere \citep{2009ARA&A..47..481A}.}
 \begin{tabular}{c c}
 \hline
 \hline
 Species (X) & Abundance \\ \hline
 He/H & 0.085\\
 O/H & $4.90\times 10^{-4}$ \\ 
 C/H & $2.69\times 10^{-4}$ \\
 N/H & $6.76 \times 10^{-5}$ \\
 Mg/H & $3.98\times 10^{-5}$ \\
 Si/H & $3.24\times 10^{-5}$ \\
 Fe/H & $3.16\times 10^{-5}$ \\
 S/H & $1.32\times 10^{-5}$ \\
 \hline
 \end{tabular}
 \tablefoot{Together, this yields a metallicity of 1.6\% in our model.}
 \label{tab:solar_values}
\end{table}

\begin{table*}
\centering
\caption{Condensation temperatures and volume mixing ratios of the chemical species.}
\begin{tabular}{c c c c}
\hline\hline
Species (Y) & $T_{\text{cond}}$ {[}K{]} & $v_{\text{Y, with C}}$  & mass fraction \\ \hline
CO & 20  & 0.29 $\times$ C/H & 13.6\% \\
N$_2$ & 20  & 0.45 $\times$ N/H & 5.32\%  \\
CH$_4$ & 30 & 0.01 $\times$ C/H &  0.20\% \\
H$_2$S & 65 & 0.1 $\times$ S/H  & 0.28\% \\
CO$_2$ & 70 & 0.1 $\times$ C/H & 7.39\% \\
NH$_3$ & 90 & 0.1 $\times$ N/H & 0.59\% \\
H$_2$O & 150 & O/H - (3 $\times$ MgSiO$_3$/H + 4 $\times$ Mg$_2$SiO$_4$/H + CO/H + 2 $\times$ CO$_2$/H) & 24.9\% \\
C (carbon grains) & 350 & 0.6 $\times$ C/H & 12.1\% \\
FeS & 704 & 0.9 $\times$ S/H  & 6.53\% \\
Mg$_2$SiO$_4$ & 1354 & Mg/H - Si/H & 6.50\% \\
Fe & 1357 & Fe/H - 0.9 $\times$ S/H  & 6.90\% \\
MgSiO$_3$ & 1500 & Mg/H - 2$\times$(Mg/H - Si/H) & 15.7\% \\
\hline
\end{tabular}
\tablefoot{Molecule condensation temperatures were taken from \citet{2003ApJ...591.1220L}. Volume mixing ratios $v_{\rm Y}$ (i.e. by number) were adopted for the species as a function of disc elemental abundances. We expand here on the different mixing ratios from \citet{2020A&A...633A..10B}.}
\label{tab:comp}
\end{table*}

\begin{table*}
 \centering
 \caption{Disc and stellar parameters of our simulations.}
 \begin{tabular}{c c c c c c c}
 \hline
 \hline
 Stellar mass [M$_\odot$] & 1.4 & 1.0 & 0.7 & 0.5 & 0.3 & 0.1 \\ \hline
 Disc mass $M_0$ [M$_\odot$]& 0.14, 0.1, 0.07, 0.05 & 0.1, 0.07, 0.05 & 0.07, 0.05, 0.03 & 0.05 & 0.03 & 0.01 \\
 Disc size $R_0$ [AU] & 320, 250, 195, 155, 110, 50 & 320-50 & 320-50 & 320-50 & 110 & 50 \\
 $\alpha$ &  $10^{-3}$, $3 \times 10^{-4}$, $10^{-4}$ & & & & &\\
 \hline
 \end{tabular}
  \tablefoot{We used the same range of $\alpha$ values for all our simulations. The planetary starting positions are 0.5, 0.75, 1.0, 1.25, 1.5, 1.75, 2.0, 2.25, 2.5, 2.75, 3.0, 3.5, 4.0, 4.5, 5.0, 6.0, 7.0, 8.0, 9.0, 10.0, 12.0, 14.0, 16.0, 18.0, and 20.0 AU. We probed six disc sizes for the 1.4, 1.0, 0.7, and 0.5 M$_\odot$ stars.}
 \label{tab:parameters}
\end{table*}

In Table~\ref{tab:parameters} we list all our disc and stellar parameters. We probed stellar masses ranging from 0.1 to 1.4 solar masses, assuming a the same metallicity, regardless of stellar type, as motivated by observations (e.g. \citealt{2022ApJ...927..123S}). We assumed that the disc mass corresponds to 10, $\sim 7$, and $\sim 5$\% of the stellar mass. We did not probe lower disc mass fractions, as we are interested in the formation of giant planets, which require disc masses of the order of a few percent of the stellar mass to grow efficiently \citep{2023A&A...679A..42S}. Although \citet{2026A&A...707A.276G} show that disc mass does not significantly affect the heavy-element content of growing planets, we probed varying disc masses to investigate whether the heavy-element content of giant planets forming in discs of the same total mass -- but around stars of different masses and disc sizes -- changes.

Discs with larger sizes can maintain a pebble flux for a longer time, with significant consequences for vapour-accreting planets (e.g. ~\citealt{2023A&A...679A..11B}). Since analyses of continuum (e.g. \citealt{2020ApJ...895..126H}) and CO disc sizes (e.g. \citealt{2022ApJ...931....6L}) do not find a statistically robust correlation between disc size and stellar mass, we probed six different disc sizes (see Table~\ref{tab:parameters}) for all stellar masses. These correspond to the disc sizes used in our previous study \citep{2023arXiv230815128M}, which were motivated by observations of protoplanetary discs \citep{2018ApJ...865..157A}. The disc size, $R_0$, in our simulations corresponds to the initial orbital position where the power law of the disc's gas surface density profile transitions into an exponential decay (e.g. \citealt{2021A&A...654A..71S}). The initial disc profile was calculated via the initial disc mass $M_0$ and is given as \citep{2017MNRAS.472.4700L}
\begin{equation}
 \Sigma_{\rm gas}(r,t) = \frac{M_0}{2\pi R_0^2} (2-\psi) \left(\frac{r}{R_0}\right)^{-\psi} \xi^{\frac{5/2-\psi}{2-\psi}} \exp \left(- \frac{(r/R_0)^{2-\psi}}{\xi}\right) \ ,
\end{equation}
where $\nu$ is the disc's viscosity and $t$ is the time. Here, $\psi=\left(\frac{{\rm d} {\rm \ln} \nu}{{\rm d}{\rm \ln}} \right)_{\rm r=r_{\rm in}} \approx 1.08$, which is evaluated at the inner edge of the disc ($r_{\rm in}=0.1$AU), and $\xi = 1 + \frac{t}{t_\nu}$, where the viscous time is given by
\begin{equation}
 t_\nu = \frac{R_0^2}{3(2-\psi)^2 \nu R_0} \ .
\end{equation}

The adopted disc sizes here are on the larger side, where millimetre-observations of class-II discs indicate mean sizes smaller than 50 AU (e.g. \citealt{2020ApJ...895..126H}). In contrast, the CO disc radius can be higher by a factor of a few (e.g. \citealt{2022ApJ...931....6L}). The positions of our CO evaporation fronts align with these constraints. In addition, class-II discs have already evolved; planet formation may have already occurred \citep{2025A&A...693A.302S}; and the millimetre-disc sizes have reduced over time (e.g. \citealt{2023A&A...673A.139A, 2025A&A...694A.311A}). As the most massive discs more efficiently form giant planets \citep{2023A&A...679A..42S}, we focused our study on these objects. As the occurrence rate of giant planets is only around 10-15\% (e.g. \citealt{J2010, 2020MNRAS.492..377W}), we can assume that giant planets form in more massive, larger discs. Observationally, discs with millimetre-sizes above 100-150 AU exist (e.g. \citealt{2018ApJ...869L..41A, 2018ApJ...869L..42H, 2020ApJ...890..130T}). We thus focused on disc with these sizes.

We used a constant grain fragmentation velocity of 5m/s, in agreement with our previous studies \citep{2023arXiv230815128M}. \citet{2026A&A...707A.276G} show that the influence of the grain fragmentation velocity on the planetary heavy-element content is not very high. In contrast, \citet{2026A&A...707A.276G} show that the heavy-element content of giant planets depends crucially on the disc's viscosity and the initial starting position of the planets. We thus used three different $\alpha$ viscosity values (see Table~\ref{tab:parameters}) and probed 25 different initial planetary positions for all stellar masses and viscosities. This allowed us to capture planets forming in different disc regions that remain exterior to the water evaporation front. Interior to this front, the heavy-element content is highest due to the evaporation of inward-drifting water-ice-rich pebbles \citep{2021A&A...654A..71S, 2024A&A...686L..17M}.

We compared the heavy-element mass fraction of the planets in our simulations with the heavy-element mass fraction derived from interior models of planets with precise mass and radius measurements \citep{2016ApJ...831...64T, 2023MNRAS.tmp.1812B, 2025A&A...693L...4M, 2025ApJ...994...43C}. While many authors have derived scaling laws for the heavy-element content of planets as a function of planetary mass, they all show the trend that the heavy-element mass increases with planetary mass. The trend of increasing heavy-element mass with planetary mass also holds for different assumed interior structures of the planet \citep{2026A&A...706A..50P}. Here, we choose to mainly compare our simulations to the scaling found in \citet{2025ApJ...994...43C}, due to their larger observational sample.

\section{Disc enrichment}
\label{sec:disc}

\begin{figure*}
 \centering
 \includegraphics[scale=0.375]{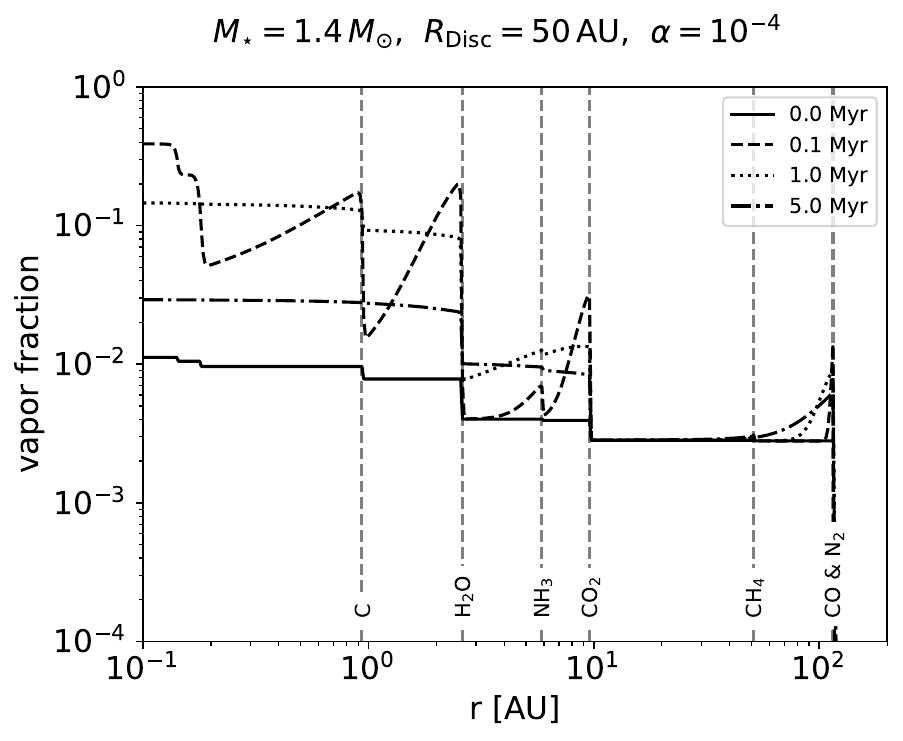} \quad
 \includegraphics[scale=0.375]{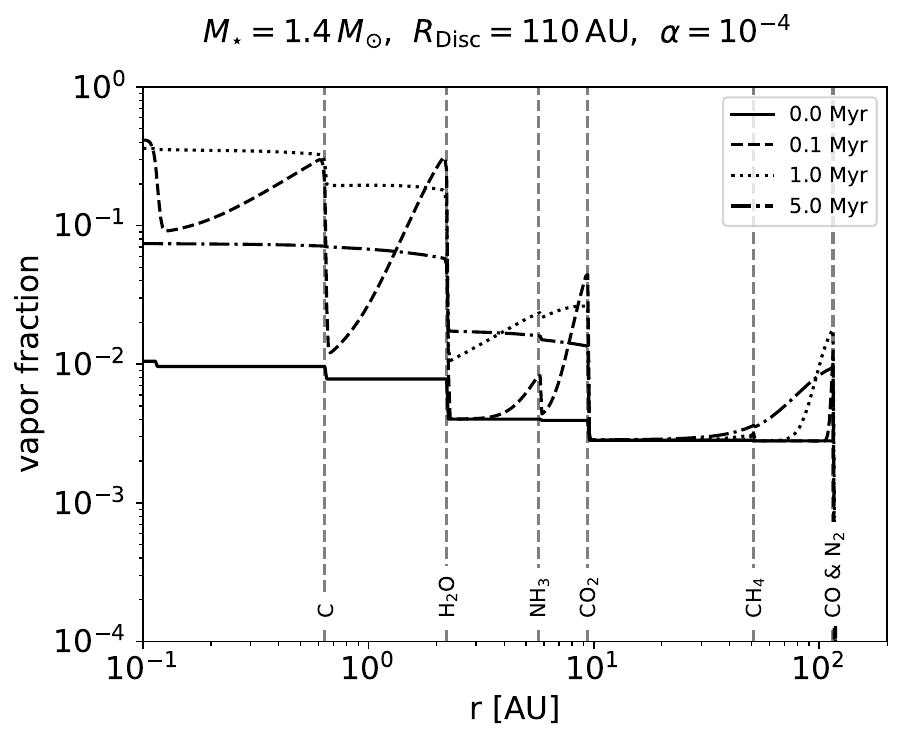} \quad
 \includegraphics[scale=0.375]{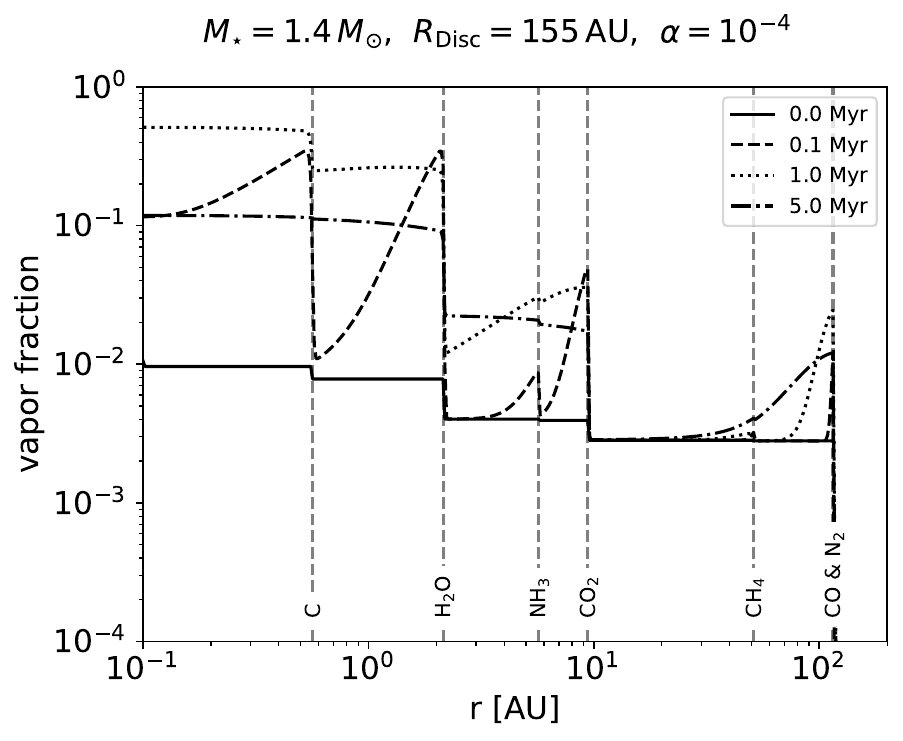} \quad
 \includegraphics[scale=0.375]{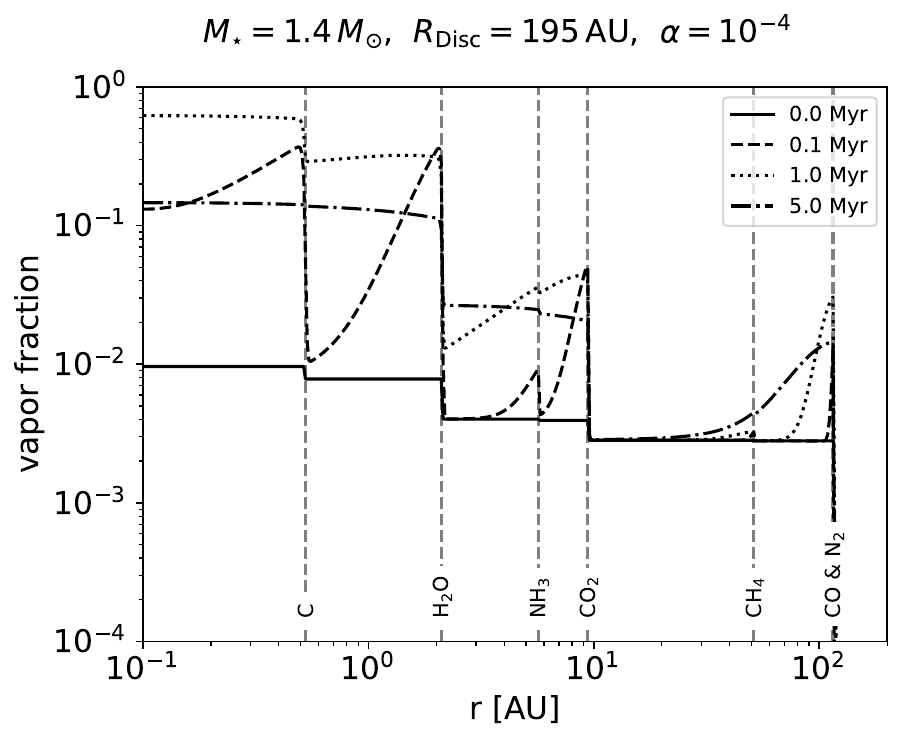} \quad
 \includegraphics[scale=0.375]{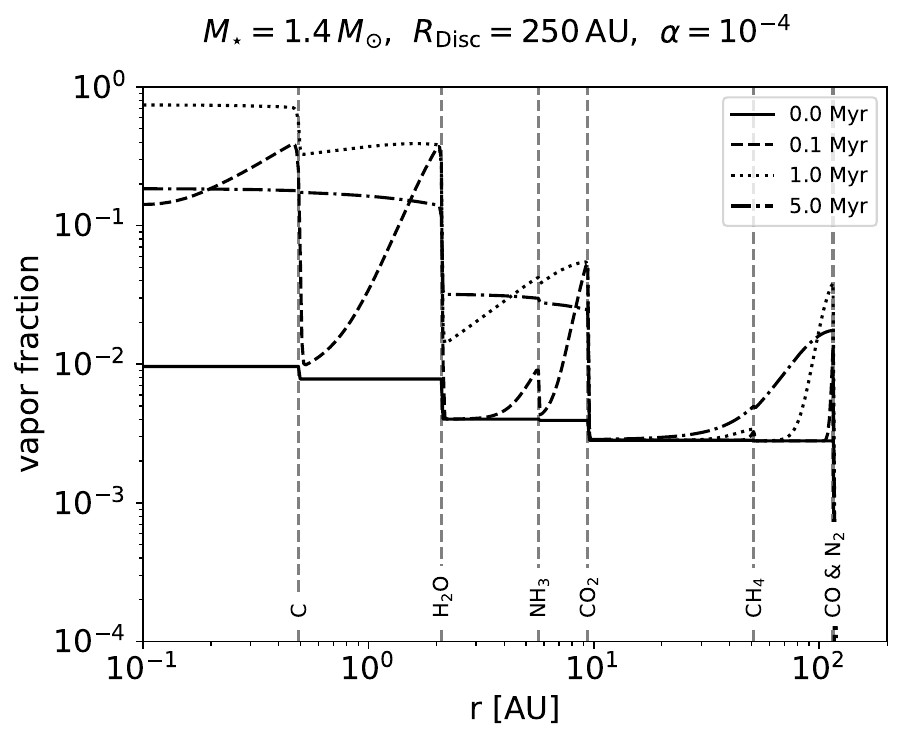} \quad
 \includegraphics[scale=0.375]{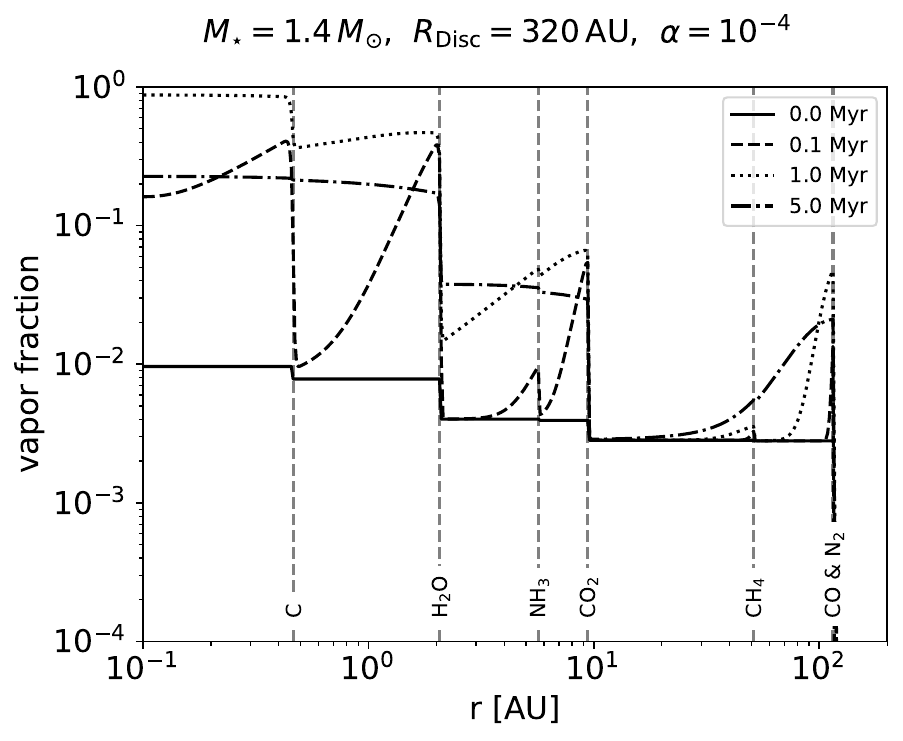}
 \caption{Vapour fraction within protoplanetary discs of different disc sizes as a function of time around a 1.4M$_\odot$ star with $\alpha=10^{-4}$. The vapour fraction is defined as the mass fraction of the sum of C, O, and N relative to the background gas, which consists of hydrogen and helium. If all solids evaporate in the inner disc, the vapour fraction at the inner edge would correspond to the initial metallicity of 0.016. The disc radii from the top left to the bottom right are 50, 110, 155, 195, 250, and 320 AU.
   \label{fig:VaporRadius}
   }
\end{figure*}

Since planets grow from material inside their natal protoplanetary discs, we first investigated how the heavy-element content in the vapour phase evolves in discs around stars with varying masses. We thus show in Fig.~\ref{fig:VaporRadius} the vapour fraction in the gas phase, which is the sum of the volatile species with evaporation temperatures below FeS (see Table~\ref{tab:comp}). This means FeS does not contribute to the vapour fraction\footnote{We do not include H$_2$S in this analysis due to its small contribution to the total heavy-element content; see Table~\ref{tab:solar_values}.}.

Figure~\ref{fig:VaporRadius} shows the time evolution of the vapour content in the gas phase for six different simulations with a stellar mass of 1.4 $M_\odot$ and six different disc sizes, assuming $\alpha=10^{-4}$, up to the fixed disc lifetime of 5 Myr. As soon as the pebbles begin to drift inwards (e.g. at 0.1 Myr), we observe high enrichment interior to the various evaporation fronts due to the evaporation of the inward-drifting pebbles. This enrichment is highest in the very inner disc regions, since all volatile species there have evaporated. This is mainly driven by the evaporation of water-ice-rich pebbles, which contribute most to the heavy-element content of the disc. Over time this high vapour enrichment, originally located only near the evaporation fronts, moves inwards, and the discs become evenly enriched interior to the various evaporation fronts (e.g. at 1.0 Myr). As time progresses further, all pebbles drift inwards, thereby reducing the vapour content of the disc as the vapour accretes viscously onto the star (e.g. at 5.0 Myr).

Figure~\ref{fig:VaporRadius}, however, reveals another trend in line with the results of \citet{2023arXiv230815128M}: the vapour enrichment interior to the water ice line takes less time in discs with smaller sizes. This evolution arises from two slightly competing effects. Discs with smaller sizes (but with the same total mass) harbour a higher surface density and are thus hotter compared to larger discs. Consequently, the inner evaporation fronts (carbon grains, water, NH$_3$, and H$_2$O) are slightly farther from the central star. On the other hand, larger discs can sustain a large pebble flux for a longer time, because the micrometre-sized dust grains take longer to grow in the outer disc regions. Once the pebble flux decays, overall vapour enrichment decreases, because the influx of volatile-rich material stops. As this occurs earlier in small discs, their vapour content decreases faster than in large discs. While initial vapour enrichment can easily reach a few tens of percent of the total amount of gas, it essentially decreases to the stellar value within a few megayears around small discs. In contrast, vapour can persist for longer times around discs with sizes of 100 AU and larger. In such cases, the vapour fraction can remain at a level of $\sim$10-20\% even at 5 Myr.

The duration of vapour enrichment in the inner disc regions is also related to the viscous evolution of the disc, as viscosity determines how long the vapour released from the inward-drifting pebbles remains in the disc. We show in Appendix~\ref{ap:discs} the vapour evolution of protoplanetary discs with varying stellar mass, disc sizes, and viscosities. The reported trend -- that discs with larger sizes can support higher vapour enrichment for a longer time  -- still holds. However, at higher viscosities, the faster viscous evolution, combined with the smaller pebble sizes due to the fragmentation-limited regime (see \citet{2012A&A...539A.148B}), results in a faster depletion of the inner disc's vapour content.

Our simulations thus clearly show that the vapour enrichment of inner discs strongly depends on disc size, in agreement with our previous simulations \citep{2023arXiv230815128M}. However, recent observations with JWST show tentative evidence that the inner disc water content may be anti-correlated with disc size and that smaller discs exhibit higher water content (e.g. \citealt{2020ApJ...903..124B, 2023arXiv230703846B, 2025A&A...694A.147G}). In contrast, \citet{2025ApJ...990L..72K} find a correlation between the water line flux and the location of the innermost dust gap detected in ALMA observations. This suggests that the inner disc water content may not be influenced by disc size but rather by pebble traps in the disc, as shown by models (e.g. \citealt{2023arXiv230701789K, 2024A&A...686L..17M, 2024ApJ...977...21E}). While outer gaps might block a fraction of the pebbles, their influence on the inner disc composition depends strongly on their depth as well as the stage of disc evolution at which they appear..

We now investigate how this effect influences the heavy-element content in giant planets in discs without outer gaps, allowing us to probe the maximal heavy-element content that giant planets can achieve in this scenario. We discuss the influence of outer gaps in Appendix~\ref{ap:gaps}.

\section{Planetary compositions}
\label{sec:composition}

In this section we investigate the growth and heavy-element content of individual planets before discussing population level trends in Sect.~\ref{sec:population}. In our work, we define the heavy-element mass of the planet as the sum of heavy-elements in the core and in the atmosphere.

\subsection{Influence of the planetary position}

\begin{figure*}
 \centering
 \includegraphics[scale=0.6]{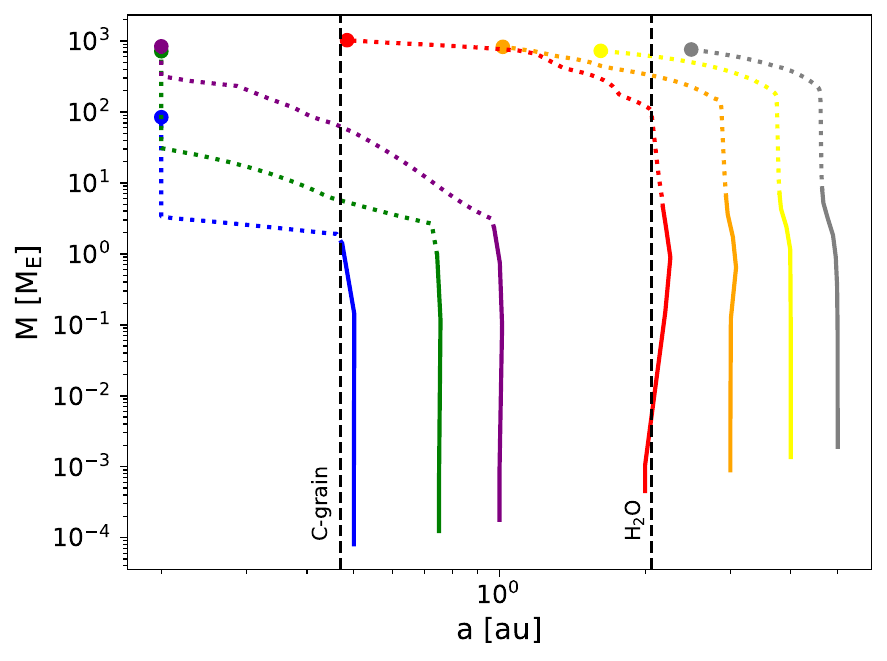} \quad
 \includegraphics[scale=0.6]{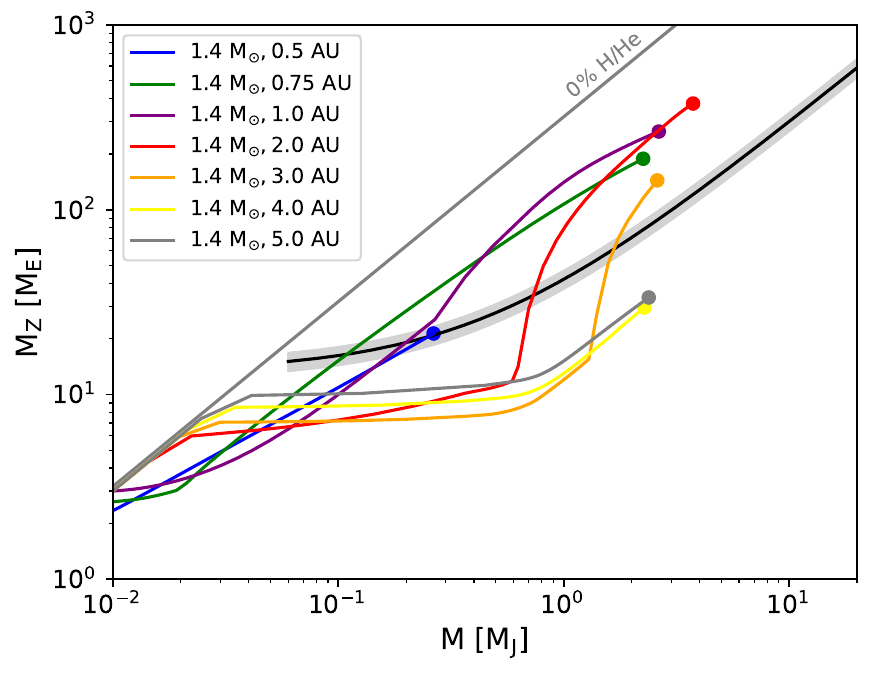}
 \caption{Left: Growth tracks of the planets starting at different positions around a 1.4 M$_\odot$ star in discs with $\alpha=10^{-4}$ and an initial disc size of $R_0=320$ AU. Here, the solid lines correspond to solid accretion, while the dashed lines correspond to gas accretion. The vertical black line marks the position of the carbon evaporation front and, farther out, the water ice line. Right: Heavy-element mass evolution of the same planets, where the black line with the grey error bars corresponds to the derived heavy-element content from \citet{2025ApJ...994...43C}. The planets forming in the inner disc regions (interior to the water ice line) begin to accrete vapour-enriched gas from the moment they accrete their envelope; thus, their heavy-element content follows a nearly horizontal line. The planets forming farther out, must first migrate interior to the water ice line to significantly enhance their heavy-element content.
   \label{fig:tracksdistance}
   }
\end{figure*}

We show in Fig.~\ref{fig:tracksdistance} the growth tracks of planets forming at seven different orbital distances up to 5 AU in discs with $\alpha=10^{-4}$ around a 1.4 M$_\odot$ star. The planets first accrete inward-drifting pebbles until they reach their pebble isolation mass. During this growth process, the planet only contains very little hydrogen (mostly due to the accretion of water ice), resulting in a heavy-element mass that is basically equal to the total planetary mass.

Once the planet begins to accrete gas, which consists of hydrogen and helium as well as a high fraction of heavy elements (see Fig.~\ref{fig:Vapor}), the planetary heavy-element content reduces by varying amounts, depending on the planet's position. The planets forming interior to the water ice line exhibit a very consistent steep increase in heavy-element content with planetary mass due to efficient accretion of water vapour. As these planets migrate -- even across the carbon grain evaporation front -- their heavy-element content increases further. These planets consistently exhibit a heavy-element content that is higher than the average derived from the interior models of \citet{2025ApJ...994...43C} once they reach around 0.3 Jupiter masses.

The planets forming exterior to the water ice line only accrete small amounts of vapour with gas (e.g. CO$_2$), resulting in a first nearly horizontal evolution in the heavy element--planetary mass diagram after reaching pebble isolation mass. However, once these planets cross interior to the water evaporation front, their heavy-element content significantly increases. This very rapid increase in the heavy-element fraction of these planets is also aided by the planet's ability to trap pebbles: as the planet reaches its pebble isolation mass, it traps a large amount of pebbles exterior to its orbit. As the planet migrates inwards, these pebbles follow and eventually release their volatiles into the gas phase when the planet crosses an evaporation front. This leads to a burst of vapour in the inner disc (see \citealt{2024A&A...691A..50E}), thereby allowing the planet to accrete large amounts of vapour and significantly increasing its heavy-element mass.

We note that planetesimal formation in the pressure perturbations generated by the planets would decrease the available pebble mass for evaporation, resulting in lower enrichment of planets migrating into the inner disc regions. However, not all pebbles are transformed into planetesimals, and millimetre observations indicate that these rings could contain up to 100 Earth mass of pebbles \citep{2018ApJ...869L..46D}, in line with our simulations. If the formation of planetesimals in the outer disc were efficient, it would increase the difference in the heavy-element mass between planets forming in the inner and outer disc regions.

Consequently, the heavy-element content of the planets shows a trend with orbital distance. Planets forming completely exterior to the water ice line show a reduced heavy-element content\footnote{The planet forming at 4 AU crosses the water ice line very late in its evolution, so that the effect of the accretion of water-rich vapour does not significantly alter its composition.} compared to planets accreting gas interior to the water ice line  (see also \citealt{2017MNRAS.469.3994B, 2021A&A...654A..71S, 2026PASJ...78..493O}).

\subsection{Influence of the disc size and viscosity}

\begin{figure*}
 \centering
 \includegraphics[scale=0.6]{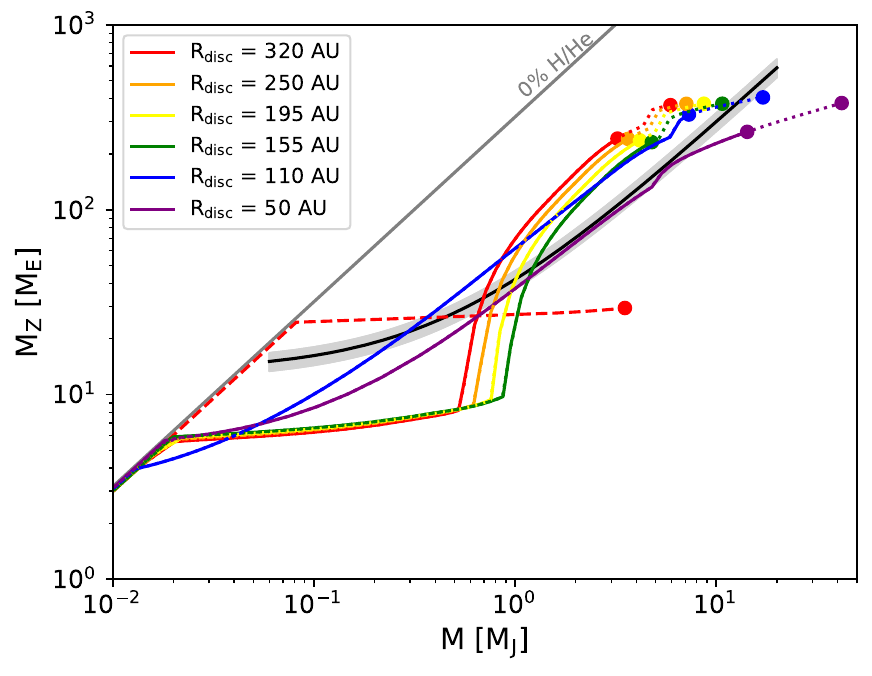} \quad
 \includegraphics[scale=0.6]{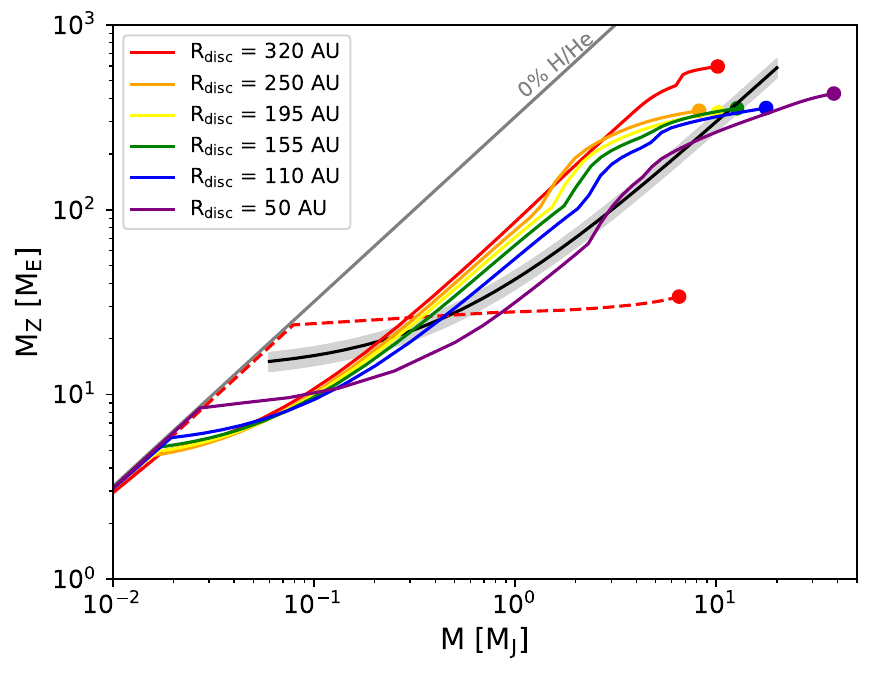}
 \caption{Evolution of the heavy-element mass of planets starting at 2.0 AU in discs with $\alpha=10^{-4}$ (left) and $\alpha=3 \times 10^{-4}$ (right) around stars with different disc sizes, up to an evolution of 5 Myr. The dotted lines in the left-hand plot correspond to planets that are evolved for 10 Myr in the disc. The dashed lines correspond to planets starting in the outer disc at 20 AU, where the pebble isolation mass is higher but the vapour enrichment is lower (Fig.~\ref{fig:Vapor}). The final planetary, heavy-element mass increases with decreasing initial disc size (for planets of the same mass).
   \label{fig:tracks}
   }
\end{figure*}

We show in Fig.~\ref{fig:tracks} the evolution of the heavy-element mass of the planets forming at 2 AU around stars of 1.4 M$_\odot$, for different disc sizes and for two different viscosities. All disc masses correspond to 10\% of the stellar mass, and all planets start at 2 AU.

All planets form exterior to the water evaporation front, and the planetary heavy-element mass is thus initially determined primarily by the pebble isolation mass. Once the planets cross the water evaporation front, their heavy-element content increases significantly (see above). We note that planets forming around stars with smaller disc sizes cross the water ice line earlier in their evolution, because the water ice line is slightly farther away from the central star.

The simulations show a clear trend: for planets above $\sim$0.7 Jupiter masses, the heavy-element mass is higher at the same planetary mass for those forming around stars with larger disc sizes. This is due to the evolution of the vapour content in the protoplanetary discs (see Fig.~\ref{fig:VaporRadius}) around stars of different disc sizes. As the vapour content in the gas phase of discs with smaller sizes decreases more rapidly, the planets forming in these disc accrete fewer heavy elements. This results in a lower overall heavy-element mass (at the same planetary mass) for these planets\footnote{Initially, the heavy-element mass of the planets forming in discs of 50 and 110 AU in size is large due to the earlier crossing of the water ice line.}.

A longer evolution time of 10 Myr results in further planetary growth as well as an increase in heavy-element mass. In this case, the rapid decrease in the heavy-element vapour content around stars with smaller disc sizes results in a slower increase in the heavy-element mass of these planets compared to their counterparts forming around stars with larger disc sizes. The planets forming around the stars with discs larger than 110 AU show a steep increase in their heavy-element content after 5 Myr. This is caused by their crossing of the carbon evaporation front, allowing them to accrete large quantities of carbon vapour as well. This process is again aided by the fact that the planet carries a large amount of pebbles in its pressure bump, which evaporate once the carbon evaporation front is crossed.

A higher viscosity ($\alpha=3 \times 10^{-4}$) of the protoplanetary disc results in a faster depletion of the vapour content in the gas phase (Fig.~\ref{fig:Vapor1e-3}). Planets forming in these discs should thus accrete lower heavy-element mass. The right panel in Fig.~\ref{fig:tracks} shows the evolution of planets starting at 2 AU in discs with $\alpha=3 \times 10^{-4}$ around stars with different disc sizes. Due to the higher viscosity, the planets begin to open gaps later and thus remain for a longer period of time in the fast type-I migration regime, resulting in a very early crossing of the water and carbon grain evaporation front for these planets. Consequently, the heavy-element mass is initially (for masses below 1 Jupiter mass) higher compared to their counterparts forming in discs with lower viscosities. However, due to the faster depletion of the vapour phase in the disc, the heavy-element mass evolution decreases towards the end of the simulations (shown by a decreased slope in the heavy-element mass--planetary mass diagram). Gas accretion in our model becomes limited by the disc's accretion rate (proportional to the disc's viscosity) once the planet opens up a gap, resulting in very high planetary masses (even above 10 Jupiter masses) compared to the planets forming in the low-viscosity environments. The object forming in the 50 AU disc is more likely a brown dwarf, assuming the typical mass threshold of 13 Jupiter masses, due to efficient gas accretion caused by the higher gas surface density of the small disc size (which has the same mass as the other disc). The trend of decreasing heavy-element mass in planets with decreasing disc size also applies to planets forming in discs with high viscosity.

For comparison we also show the heavy-element mass evolution of planets forming at 20 AU around a 1.4 M$_\odot$ star in a disc with a size of 320 AU. The planet remains exterior to the water evaporation front at all times, resulting in a total heavy-element mass that is nearly entirely dominated by the pebble isolation mass (shown by the horizontal evolution of the planetary heavy-element content at around 25 Earth masses). The planet forming in the disc with $\alpha=3 \times 10^{-4}$ also remains in the outer disc regions and exhibits the same evolution.

\subsection{Influence of the stellar and disc mass}

\begin{figure*}
 \centering
 \includegraphics[scale=0.6]{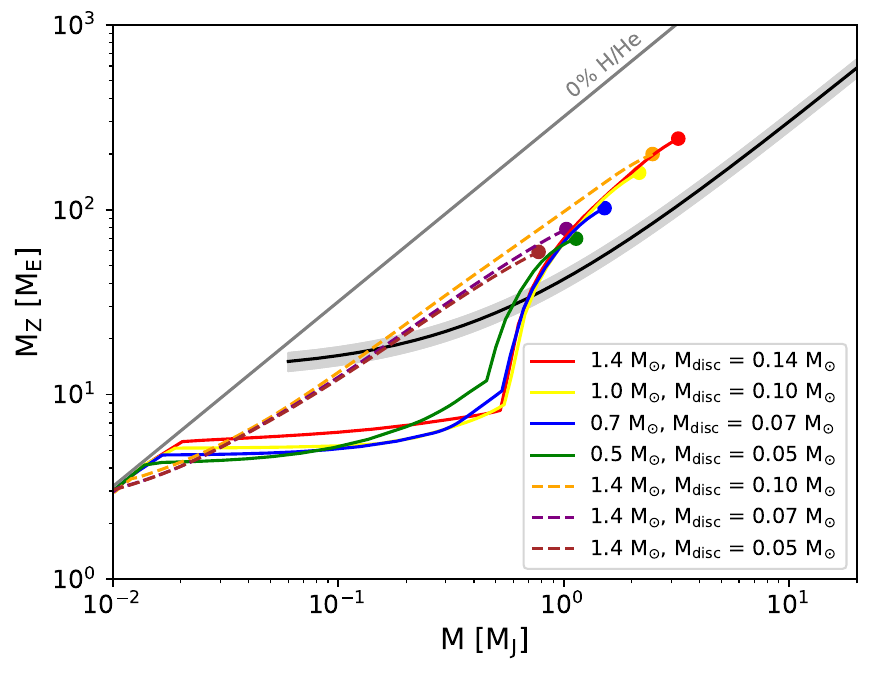} \quad
 \includegraphics[scale=0.6]{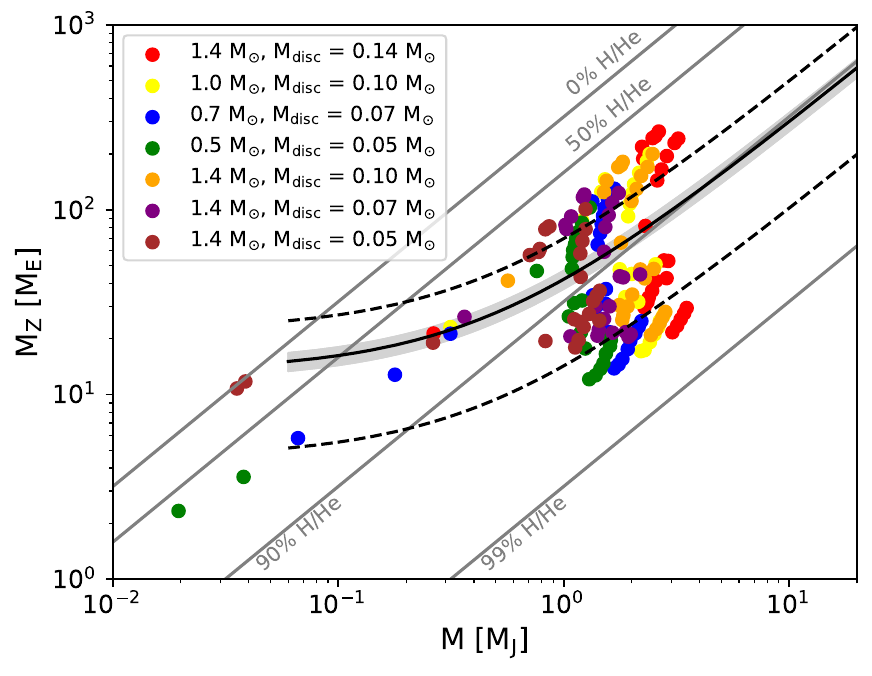}
 \caption{Left: Evolution of the heavy-element mass of planets starting at 2.0 AU in discs with $\alpha=10^{-4}$, a size of 320 AU, and different disc masses around stars of varying stellar masses. Right: Final heavy-element mass of planets starting at various initial positions in the same discs. The planets forming in discs with the same mass but lower stellar mass exhibit a slightly lower heavy-element content due to the faster decay of the gaseous heavy-element content within the disc (see Fig.~\ref{fig:Vapor}).
   \label{fig:tracksDisc}
   }
\end{figure*}

To investigate whether the heavy-element content of planets is more strongly influenced by stellar or disc mass, we show in Fig.~\ref{fig:tracksDisc} the heavy-element mass evolution of planets forming around stars with varying stellar and disc masses. The disc size is fixed at 320 AU for all simulations. The planets begin again at 2.0 AU and evolve in discs with $\alpha=10^{-4}$.

Our simulations show that the heavy-element mass of planets forming in discs with the same total mass -- but around more massive stars (i.e. where the star/disc mass fraction is lower), is generally (slightly) higher. This is caused by the (slightly) slower depletion of the vapour content around higher-mass stars (Figs.~\ref{fig:VaporRadius} and~\ref{fig:Vapor}), allowing the planets to accrete gas enriched with more vapour, which leads to a slightly higher heavy-element mass. We note that the evolution of the planets forming around the 1.4 $M_\odot$ star shows a constant increase in their heavy-element content compared to planets forming around lower-mass stars with the same disc mass. This is because the slightly lower disc mass reduces the planet's growth, increasing the time spent in type-I migration. This, in turn, enables an earlier crossing of the water ice line, which is located further out compared to the lower-mass stars.

We show in the right part of Fig.~\ref{fig:tracksDisc} the final heavy-element mass of planets forming at initial orbital distances ranging from 0.5 to 20.0 AU around the same stars (as in the left part of the figure) and with the same disc properties. Also, on a population level, planets forming around higher-mass stars appear be slightly more enriched than their those forming around lower-mass stars. This is caused by the slightly slower decay of the vapour content of disc's gas phase around higher-mass stars (see Appendix~\ref{ap:discs}). Combined with our previous results, this shows that the heavy-element content of giant planets is more affected by disc size than stellar mass.

\section{Planetary population}
\label{sec:population}

In this section we analyse the general outcome of our large set of simulations (see Table~\ref{tab:parameters}). Generally, planets forming in discs with higher viscosities have higher masses due to the higher disc accretion rate, which sets the planetary gas accretion rate. We note that a shorter disc lifetime would result in less massive planets (see Appendix~\ref{ap:lifetime}).

\subsection{General trends}

\begin{figure*}
 \centering
 \includegraphics[scale=0.6]{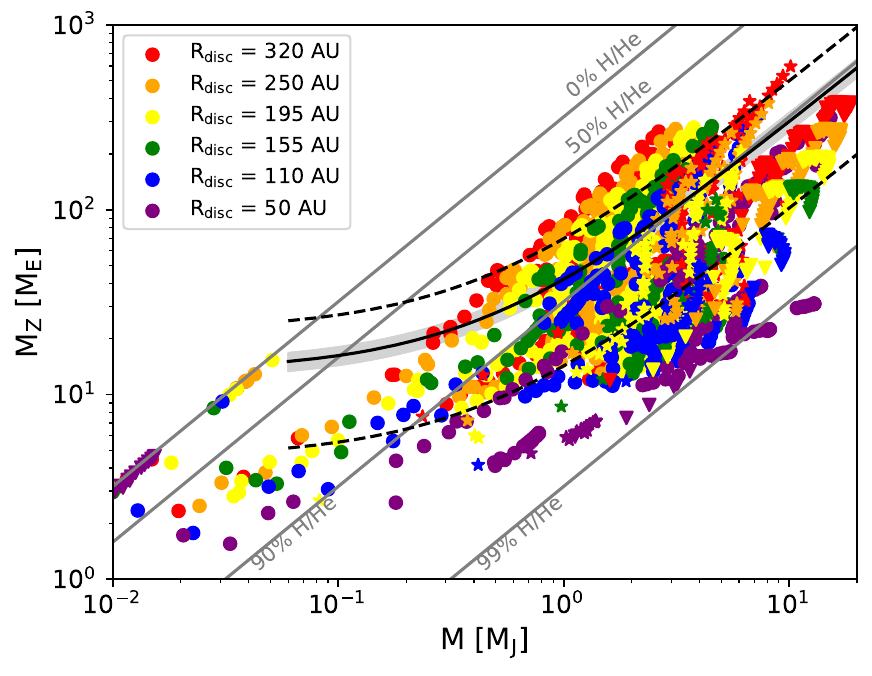} \quad
 \includegraphics[scale=0.6]{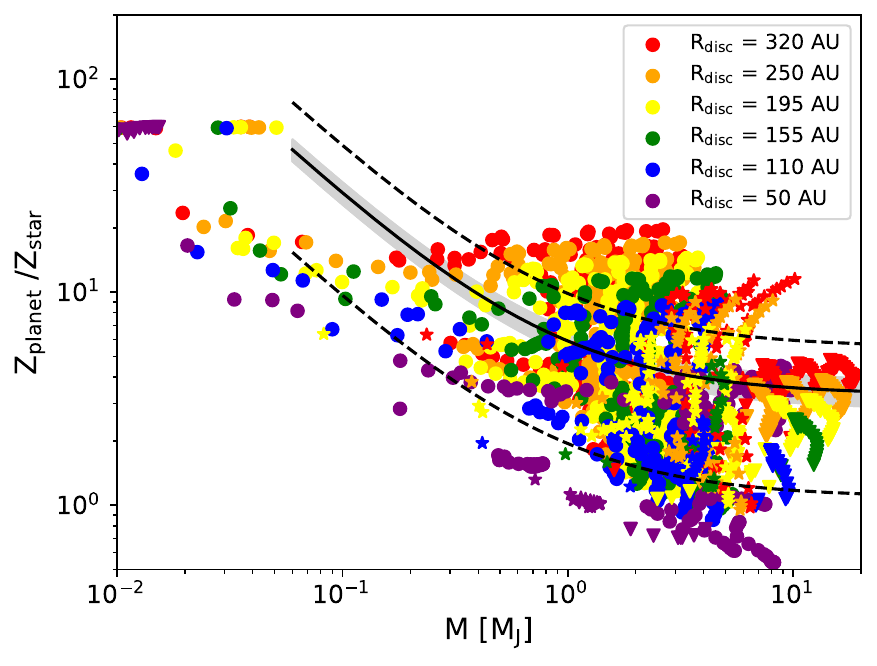}
 \caption{Heavy-element mass (left) and metallicity (right) of the simulated planets around different stellar types with varying disc sizes as a function of planetary mass. The different disc sizes are indicated by different colours. The black line with the grey error band corresponds to the heavy-element-mass fit from \citet{2025ApJ...994...43C}, where the dashed black lines correspond to the 1$\sigma$ error. The circles correspond to simulations with $\alpha=10^{-4}$; the stars correspond to $\alpha=3 \times 10^{-4}$; and the upside down triangles correspond to $\alpha=10^{-4}$.
   \label{fig:MZplanet}
   }
\end{figure*}

We show in Fig.~\ref{fig:MZplanet} the final heavy-element mass as well as the metallicity of the planets divided by the stellar metallicity (Z=0.016) from all our simulations. The different colours correspond to the different disc sizes, while the different symbols correspond to different $\alpha$ values.

Figure~\ref{fig:MZplanet} verifies the trend of the previously discussed heavy-element mass evolution of individual planets: planets forming around stars with larger disc sizes can harbour higher heavy-element content compared to planets of the same mass that form around stars with smaller disc sizes. This trend holds (individually) for all three different viscosities investigated here. As discussed above, we generally see that giant planets forming in discs with higher viscosities attain higher overall mass and thus contain more heavy elements. However, these planets have a higher H/He fraction. Our simulations also confirm the trend that the overall heavy-element mass increases with planetary mass.

Planets forming around stars with larger disc sizes can cover the whole range of the observationally derived heavy-element content of giant planets. On the other hand, planets forming around stars with smaller disc sizes cannot reach such high heavy-element masses. This indicates that discs with larger sizes are required to explain the full range of the heavy-element masses of giant planets.

The formation of planets above 2-3 Jupiter masses, requires higher viscosities in our model. Also, in this case, planets forming around stars with larger disc sizes harbour a higher heavy-element mass thank those forming around stars with smaller disc sizes. While the total heavy-element mass increases with increasing planetary mass, the H/He fraction decreases, in line with the observational trend.

We show in the right-hand side of Fig.~\ref{fig:MZplanet} the ratio of the planetary to stellar metallicity. The planetary metallicity is defined as the total amount of heavy elements divided by the total mass of the planet. As expected, the ratio Z$_{\rm planet}$/Z$_\star$ decreases with increasing planetary mass, because the increase in total heavy-element mass with planetary mass does not follow a 1:1 ratio (left in Fig.~\ref{fig:MZplanet}). As the planets forming around stars with smaller disc sizes harbour fewer heavy elements, their Z$_{\rm planet}$/Z$_\star$ ratio is also generally smaller and reaches lower values (at the same planetary mass) compared to the planets forming around stars with larger discs. We note that the overall heavy-element mass of the planet originates mostly from the accretion of oxygen and carbon vapour, resulting in very heavily enriched planetary atmospheres (see Appendix~\ref{ap:abu}), in line with exoplanet observations (e.g. \citealt{2025A&A...699A.342B, 2026AJ....171...31L}). On the other hand, the planets forming around stars with smaller disc sizes can show sub-stellar metallicities and abundances, in general agreement with exoplanet observations around these types of stars (e.g. \citealt{2026AJ....171..294A, 2026arXiv260316464K}).

At the low-mass end (below 0.1 Jupiter masses), our simulations show several planets with very high Z$_{\rm planet}$/Z$_\star$ ratios as well as high heavy-element content relative to their total mass. These planets do not enter a stage of runaway gas accretion and are thus dominated by their core, which contains more than 90\% of heavy elements (the remaining part may contain hydrogen due to the accretion of for example water ice).

We note that our simulations produce few planets in the mass range of 0.1 to 0.5 Jupiter masses. This is a consequence of allowing the planets to evolve for 5 Myr. Over this timescale, planets can accrete more gas from the disc, and smaller planets may eventually contract their planetary envelope and become gas giants. Shorter disc evolution times would produce lower-mass planets. Essentially, an earlier disc dissipation time would allow any position along the evolutionary lines in Figs.~\ref{fig:tracksdistance},~\ref{fig:tracks}, and~\ref{fig:tracksDisc}, covering planetary masses below 0.5 Jupiter masses. We show the heavy-element mass of the planets after 2 and 3 Myr in Appendix~\ref{ap:lifetime}, allowing us to cover the parameter space with planets below 0.5 Jupiter masses in our planet formation model.

The heavy-element content curve provided by \citet{2025ApJ...994...43C} fits the heavy-element content of our simulated planets very well. In our model, the heavy-element mass of planets at the low-mass end is determined by the core mass (set by the pebble isolation mass), while planets above 0.1 Jupiter masses are dominated by the accretion of vapour-enriched gas. However, we caution against directly comparing simulations to the derived heavy-element content from interior models and observations. In reality, planets originate from different environments (e.g. different stellar masses, metallicities, and initial orbital positions). However, the observational sample used to derive the reported trend originates from a semi-random collection of warm Jupiters that, coincidentally, have well-constrained mass and radius measurements. These measurements allow us to determine their heavy-element content via interior structure models. However, the origin of the warm Jupiter population is unclear (disc migration vs planet--planet scattering) and may influence the heavy-element content of the probed planets.

\subsection{Origin of the heavy-element content}

\begin{figure*}
 \centering
 \includegraphics[scale=0.6]{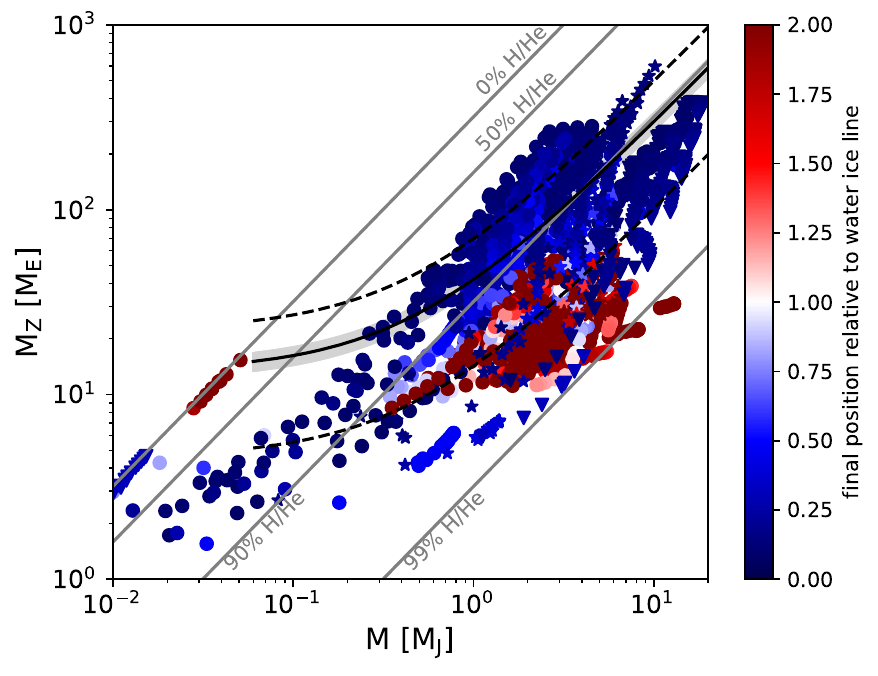} \quad
 \includegraphics[scale=0.6]{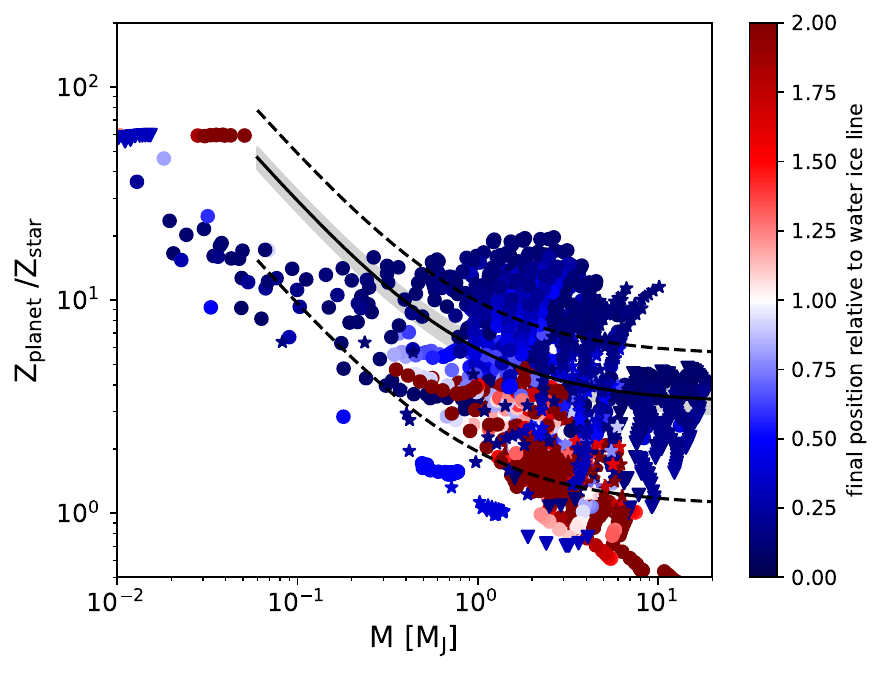}
 \caption{Heavy-element mass (left) and metallicity (right) of planets starting at different orbital distances in a disc with $\alpha=10^{-4}$ (circles), $\alpha=3 \times 10^{-4}$ (stars), and $\alpha=10^{-3}$ (triangles) colour-coded by their final orbital position relative to the water ice line in the corresponding systems. The black line with the grey error bands correspond to the fits from \citet{2025ApJ...994...43C}.
   \label{fig:MZdistance}
   }
\end{figure*}

Figure~\ref{fig:VaporRadius} shows that the vapour content in the gas phase increases towards the inner disc, particularly around the water ice line. Planets migrating across this line, therefore, should harbour high heavy-element content (Fig.~\ref{fig:tracksdistance}). We show in Fig.~\ref{fig:MZdistance} the heavy-element mass and the Z$_{\rm planet}$/Z$_\star$ ratios of the planets as functions of their total mass and final orbital position colour-coded by their relative position to the water ice line. As the position of the water ice line is a strong function of the stellar mass and disc viscosity (e.g. Figs.~\ref{fig:Vapor} and~\ref{fig:Vapor1e-3}), we used the position relative to the water ice line rather than absolute distances in AU. We remind the reader that the temperature structure of our disc does not evolve over time.

Figure~\ref{fig:MZdistance} shows a very clear trend of the total heavy-element mass of the planets as a function of their distance to the water ice line. All planets with more than $\sim$50 Earth masses of heavy elements have either formed completely or migrated interior to the water ice line. In contrast, planets below $\sim$50 Earth masses could remain exterior to the water ice line but may have also migrated interior to it. The inward-migrating planets with these low heavy-element masses formed predominantly around stars with small disc sizes and in discs with high viscosity. Both -- small disc sizes and high viscosities -- reduce the vapour mass fraction in the inner disc (Figs.~\ref{fig:Vapor} and~\ref{fig:Vapor1e-3}), leading to the formation of planets with lower heavy-element content. The trend found here -- that planets with high heavy-element content have accreted gas interior to the water ice line -- is similar to the study of \citet{2026PASJ...78..493O}. However, our study shows that planets interior to the water ice line may also exhibit low heavy-element content. Conversely, our study shows that planets forming exterior to the water ice line do not have high heavy-element content.

The Z$_{\rm planet}$/Z$_\star$ ratio reflects this trend as well. Giant planets with a Z$_{\rm planet}$/Z$_\star$ ratio greater than $\approx$5 migrated into the inner disc and accreted large amounts of vapour. In contrast, planets with a Z$_{\rm planet}$/Z$_\star$ ratio lower than 5 may have either formed completely exterior to the water ice line or migrated across it if they formed around low-mass stars or in high-viscosity environments.

We note that the planets following the 0\% H/He line or exhibiting a Z$_{\rm planet}$/Z$_\star$ ratio of around 60, are planets that did not accrete gas efficiently. These planets are dominated by their core and only contain very little hydrogen, due to the accretion of water ice into their core, regardless of their final orbital position.

We note that all planets have a heavy-element fraction that is essentially higher than the stellar value. This is due to the formation of the planetary core, which can already contain a few Earth masses of heavy elements, regardless of how efficient vapour accretion is. Consequently, these planets show total heavy-element content below the mean value derived in \citet{2025ApJ...994...43C}.

\subsection{Trends with stellar metallicity}

Our standard simulations use Z=0.016, which corresponds to the solar metallicity (Table~\ref{tab:solar_values}). To probe how a change in the host star's metallicity affects the planetary heavy-element content, we also probed slightly higher and slightly lower metallicities, corresponding to Z=0.010 and Z=0.024, respectively. In the bracket notation, these changes correspond to
\[
  [Z/H] = +0.2 \ , \ [Z/H] = -0.2 \ .
\]

\begin{figure*}
 \centering
 \includegraphics[scale=0.6]{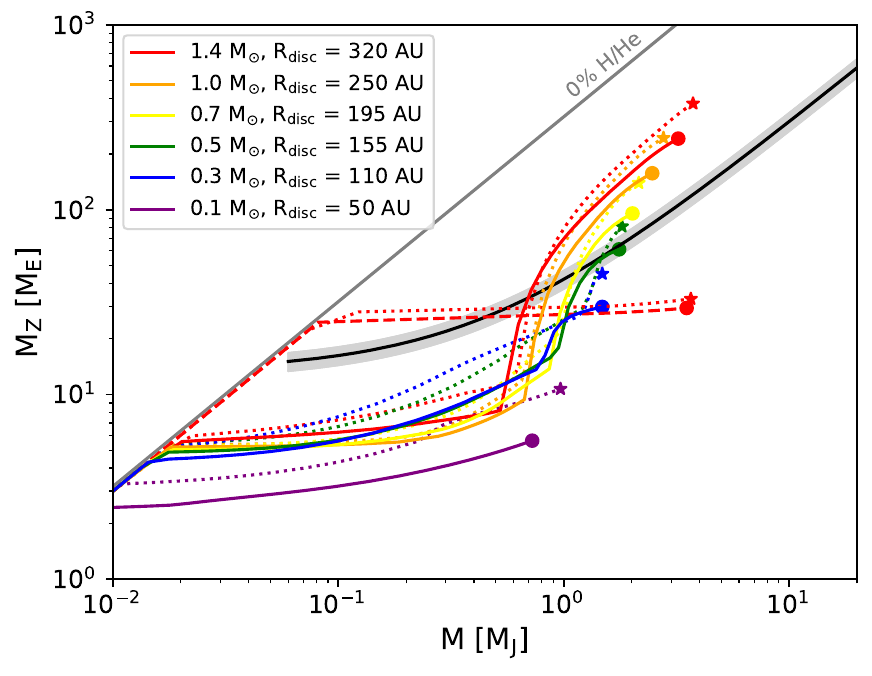} \quad
 \includegraphics[scale=0.6]{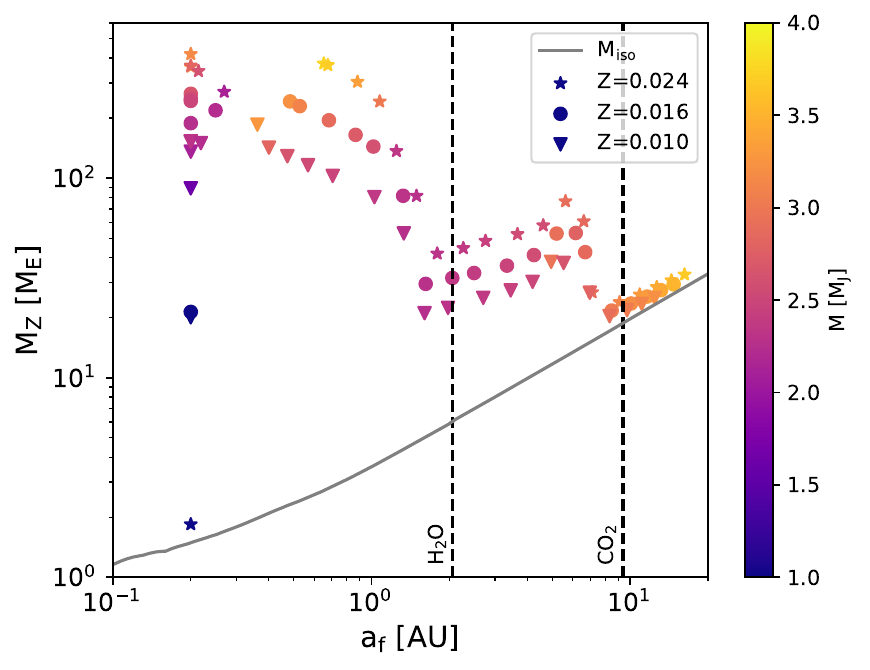}
 \caption{Left: Evolution of the heavy-element mass of planets starting at 2.0 AU in discs with $Z=0.024$ (dotted) and $\alpha=10^{-4}$, up to an evolution of 5 Myr. The solid lines correspond to planets forming in discs with a standard metallicity of $Z=0.016$. The dashed lines correspond to a planet starting in the outer disc with $Z=0.016$ at 20 AU, where the pebble isolation mass is higher. The planets start in discs with different stellar masses and disc radii. Right: Final heavy-element mass of planets forming at different locations around a 1.4 $M_\odot$ star with a disc size of 320 AU and varying metallicities. The vertical dashed line at $\approx 2$ AU corresponds to the water ice line, interior to which the heavy-element content of planets increases significantly. The second dashed line at $\approx 9.5$ AU corresponds to the CO$_2$ evaporation front, while the grey line marks the pebble isolation mass. The planets forming in the very inner disc regions around fewer metal-rich stars can harbour a higher heavy-element mass compared to planets forming exterior to the water ice line around a high-metallicity star. Exterior to the CO$_2$ evaporation front, the disc metallicity does not significantly influence the total heavy-element mass of the planets, because this mass is set by the planet's core, which is set by the pebble isolation mass that is independent of the disc metallicity.
   \label{fig:tracksZ}
   }
\end{figure*}

The heavy-element content of the planets formed in the inner disc scales roughly with the overall metallicity, as most of the C, O, and N has evaporated. The planets can then proportionally accrete more vapour, and we expect that planets forming in high-metallicity environments in the inner regions should be more enriched in heavy elements. Figure~\ref{fig:tracksZ} confirms this trend. However, we also observe a trend with orbital distance, where planets forming farther out in the disc are less affected by an increase in stellar metallicity.

The left-hand side of Fig.~\ref{fig:tracksZ} shows the evolution of the planetary heavy-element mass as a function of planetary mass. The planets forming in the discs with higher heavy-element content exhibit a slightly higher heavy-element content than those in the discs with lower metallicities, regardless of stellar mass and disc size. This is caused by the initial formation location of the planets at 2.0 AU, interior to the CO$_2$ evaporation front, where some vapour can already be accreted. However, once the planets migrate across the water ice line, the higher disc metallicity results in much higher heavy-element masses. In contrast, the planets forming at 20 AU remain exterior to the CO$_2$ evaporation front and, thus, exterior to the water evaporation front. Their heavy-element mass is determined by their core mass alone. As the pebble isolation mass is independent of the disc's metallicity, the planets essentially feature the same heavy-element mass regardless of their birth metallicity environment. Such small differences in heavy-element mass are explained by the accretion of a slightly higher fraction of CO- and nitrogen-rich gas around higher-metallicity stars.

The right-hand side of Fig.~\ref{fig:tracksZ} shows the heavy-element mass of planets formed around a 1.4 M$_\odot$ star in discs with three different metallicities and $\alpha=10^{-4}$ as functions of their final orbital position. The planets start up to 20 AU and slowly migrate inwards during their evolution. Figure~\ref{fig:tracksZ} reveals three different distinct regions relative to the final heavy-element content of the formed planets.

Exterior to the CO$_2$ evaporation front, the disc's vapour content is very low, because CO and CH$_4$ vapour are very inefficient at moving inwards (Fig.~\ref{fig:VaporRadius}) and also contain only small amounts of carbon and oxygen. The heavy-element mass of the planets is thus determined primarily by their core mass, which is set by the pebble isolation mass, with a small additional contribution from the CO, CH$_4$, and N$_2$ present in the gas phase of the disc. As the pebble isolation mass is independent of the disc's metallicity (e.g. \citealt{2018arXiv180102341B}), there is no difference in the heavy-element mass of the planets formed in discs with varying metallicities.

Between the water and CO$_2$ evaporation front, the heavy-element content of the planets becomes dominated by gas accretion, resulting in a heavy-element mass that is about a factor of 2 greater than that provided by the planetary core. As the accretion of vapour-enriched gas begins to drive the total heavy-element mass, the planets forming around stars with varying metallicities exhibit different heavy-element masses. As expected, planets forming around high-metallicity stars are more enriched in this region. The slight increase in heavy-element mass towards the CO$_2$ evaporation front is caused by two effects: (i) the larger pebble isolation mass of the planets forming farther out and (ii) the slightly higher CO$_2$ vapour content near the CO$_2$ evaporation front during the first few megayears (Fig.~\ref{fig:VaporRadius}).

Interior to the water ice line, the heavy-element content is highest due to the accretion of water and carbon vapour. The planets in these regions reach heavy-element masses of up to 400 Earth masses, with the expected trend that planets around metal-rich stars have higher heavy-element masses.

Together, our simulations reproduce the expected trend: planets around more metal-rich stars are more metal-enriched, provided that they form interior to the CO$_2$ evaporation front. However, planets forming in the outer disc, even around metal-rich stars, would harbour only very low heavy-element masses, as determined by the pebble isolation mass.

\section{Discussion}
\label{sec:discussion}

\subsection{Shortcomings of our model}

The heavy-element content of our model is driven by the accretion of vapour in the inner disc regions. Planets forming in the outer disc regions are dominated by the core mass, set by the pebble isolation mass \citep{2014A&A...572A..35L, 2018arXiv180102341B}. Our model thus ignores the formation and accretion of planetesimals as well as the collisions between the planets themselves. However, \citet{2023A&A...679L...7D} show that the formation of planetesimals in discs decreases the vapour content in the gas phase and, consequently, the heavy-element content of forming planets. Including the accretion of the formed planetesimals results in a relatively low heavy-element fraction, because the planetesimal-to-gas ratio in the discs is lower than the vapour content \citep{2023A&A...679L...7D}, as also found by \citet{2026ApJ..1007...19B}. Furthermore, the accretion of planetesimals is not 100\% efficient, whereas the accretion of gas enriched with vapour occurs at the corresponding mass fraction of the vapour within the disc. \citet{2020A&A...633A..33S} used N-body simulations to study how many planetesimals could be accreted onto migrating giant planets. However, their model failed to produce heavy-element content above 100 Earth masses, which is a typical outcome in our simulations. In our study we are mostly interested in the maximum heavy-element content the planets could harbour, which is why our study is restricted to a pure pebble accretion scenario.

\citet{2020MNRAS.498..680G} used collisions between giant planets with massive cores (up to 30 Earth mass cores) to explain the heavy-element content of giant planets. Although their model could eventually produce these high heavy-element fractions by assuming significant gas loss during the collisions, they did not model the collision via N-body interactions, which typically show less frequent collisions (e.g. \citealt{2012ApJ...751..119B, 2023A&A...674A.178B}) of these objects. Furthermore, the formation of many Jupiter-type planets in the inner disc is not reproduced in N-body simulations \citep{2021MNRAS.505.2500P}. \citet{2025ApJ...994...43C} also used a simple estimate via planetary collisions to explain the heavy-element masses of giant planets. They concluded that collisions between planets cannot explain the Jupiter-mass planets that feature heavy-element content above $\sim$50 Earth masses. However, collisions may be required to explain planets below $\sim$40 Earth masses that have heavy-element mass fractions of nearly 100\%. Our model has difficulties producing these type of planets, which would essentially be planets that have just reached the pebble isolation mass but never accreted gas (see also Appendix~\ref{ap:lifetime}).

Our chemical partitioning model shows that roughly 75\% of the heavy-element fraction is made up of O, C, and N. These elements are highly volatile, exist predominantly in the gas phase, and can therefore be accreted efficiently onto the planet with the gas. However, some of the oxygen is hidden in the refractory solids, so that at t=0 roughly 60\% of the heavy elements are in a gaseous form in the inner disc regions. If accretion thus proceeds only via planetesimals (which is not 100\% efficient), a high fraction of heavy elements will not be accreted, resulting in lower heavy-element enrichment of the planets (see \citealt{2023A&A...679L...7D}).

The chemical partitioning model (Table~\ref{tab:comp}) assumes that 60\% of the carbon is within refractory carbon grains. However, if a lower fraction of carbon were in carbon grains, and instead the fraction of CO and CO$_2$ were higher, the partitioning model would predict less water, as more oxygen would be stored in CO and CO$_2$ \citep{2020A&A...633A..10B}. Consequently, the inner disc would be less enriched in water and carbon vapour, resulting in less enriched planets. \citet{2021A&A...654A..71S} used a partitioning model without carbon grains and consequently found that planets are less enriched by vapour than in our study here. The water content would also be influenced by the stellar C/O ratio \citep{2020A&A...633A..10B, 2023arXiv230105034C}, showing the importance of using detailed stellar abundances if the formation of individual systems is explained \citep{2026A&A...706A..30O}.

Our model assumes a smooth disc enabling efficient inward drift of pebbles, which can thus enrich the inner disc regions. However, outer pressure perturbations could hinder the inward drift of pebbles (e.g. \citealt{2012A&A...538A.114P, 2023arXiv230701789K, 2024A&A...686L..17M, 2026MNRAS.548ag423L}). Depending on the timing, perturbation strength, and disc viscosity, pebbles could be efficiently blocked, reducing the inward-drifting pebble flux and consequently lowering the vapour enrichment in the inner disc (e.g. \citealt{2023arXiv230701789K, 2024A&A...686L..17M}). If pressure perturbations would be efficient in the early stages of disc evolution, our model predicts a lower heavy-element mass fraction. However, if the outer pressure perturbations were present from the beginning of the disc evolution, giant planet formation would also be quenched, as the pebble flux would be reduced significantly \citep{2019A&A...627A..83L}. We show the evolution of the planetary heavy-element content in discs with outer gaps in Appendix~\ref{ap:gaps}.

Models with multiple forming planets also show that inner planet enrichment is not altered significantly by the outer gap-opening planets, if the planets form in sequential order from inside out \citep{2024A&A...691A..50E}. On the other hand, very young discs do not show any signs of substructure (e.g. \citealt{2022A&A...661A..66Z}), indicating that, at least initially, large amounts of pebbles could make their way into the inner disc, causing the enrichment required to form giant planets with high heavy-element content. This high vapour enrichment is also observed in discs with gap and ring-like structures (e.g. \citealt{2023arXiv230703846B}), indicating that vapour accretion plays a role in shaping the giant planet population.

Our simulations are agnostic about disc size as a function of stellar mass. However, if the disc size were to decrease with decreasing stellar mass (e.g. \citealt{2010ApJ...723.1241A}), our simulations would predict that giant planets around lower-mass stars should be less enriched in heavy elements compared to planets forming around higher-mass stars. This would align with the trend identified by \citet{2025A&A...693L...4M}; see Appendix~\ref{ap:discradius}. On the other hand, the observational trend of disc size with stellar mass seems to indicate that disc sizes do not strongly depend on stellar mass (e.g. \citealt{2020ApJ...895..126H, 2022ApJ...931....6L}). This shows the importance of combining constraints from exoplanets and protoplanetary discs to understand planet formation.

\subsection{Implications for giant planet formation}

The high heavy-element content of giant planets has been used to infer their formation location \citep{2026A&A...706A..30O}. However, such studies require detailed stellar abundances as input to constrain the heavy-element content during the gas disc phase. Here, we extend the simulations to also understand the heavy-element content of giant planets around different stellar types.

The analysis of \citep{2025A&A...693L...4M} revealed that giant planets forming around M dwarfs should harbour a lower heavy-element content compared to planets of the same total mass around FGK stars. In particular, they find that the giant planets around M dwarfs harbour a maximum of 50 Earth masses of heavy elements with a mean of approximately 20 Earth masses. In contrast, the planets around FGK type stars in their sample can have heavy-element masses above 100 Earth masses. However, the study by \citet{2025ApJ...994...43C} does not recover this trend. Our simulations also only show a very small dependency of stellar mass on the heavy-element content of forming giant planets (Fig.~\ref{fig:tracksDisc}), in line with the observational trend of \citet{2025ApJ...994...43C}.

Our simulations predict that giant planets with high heavy-element content (e.g. above the fit of \citealt{2025ApJ...994...43C}) migrate inwards across the water ice line and should thus have obliquities and eccentricities close to zero. In contrast, planets with a low heavy-element content may form exterior to the water ice line. This implies that scattering events would be needed to bring these planets to their currently observed orbits. Thus, our simulations predict that planets with lower heavy-element content are either more likely to have significant obliquities or lie on an eccentric orbit compared to planets with a high heavy-element content. Together, our simulations suggest that metal-rich planets form via disc migration, while metal-poor, warm giant planets either experience scattering events or migrate inwards in high-viscosity environments.

Metal-rich stars are more prone to host giant planets (e.g. \citealt{2004A&A...415.1153S, 2005ApJ...622.1102F, J2010}). Naturally, our model predicts that planets forming around more metal-rich stars are more metal-rich, provided they form within the same evaporation fronts (Fig.~\ref{fig:tracksZ}). However, the observations from \citet{2019AJ....158..239T} indicate that more metal-rich stars do not, on average, produce more metal-rich planets. Within our simulations this observation can be naturally explained by different formation locations. If planets form exterior to the water ice line, even around metal-rich stars, they will exhibit lower heavy-element mass compared to planets forming interior to the water ice line, even if the star is metal-poor. In particular, we find that the heavy-element mass of planets can increase by a factor of 10 from exterior to interior to the water ice line, while the heavy-element content seems to scale just linearly with the host star's metallicity (Fig.~\ref{fig:tracksZ}). Planet formation itself, on the other hand, is more efficient at greater orbital distances around metal-rich stars (e.g. \citealt{2015A&A...582A.112B, 2021A&A...656A..69E}), naturally leading to the large variety of heavy-element masses in giant planets forming around such stars.

In our model, the inward-migrating planets undergo an interesting phase during their formation: as the planets migrate into regions with increasingly higher vapour content, the heavy-element content in the planetary atmosphere increases as a function of planetary radius (assuming that there is no mixing). This could have potential implications for the observations of planetary atmospheres and how these link to the bulk abundances (see \citealt{2024ApJ...977..227K}).

Recently, \citet{2026arXiv260526216L} highlighted a potential problem: the high fraction of free-floating planets should contain a higher fraction of heavy elements compared to the amount available during the disc stage. However, their estimates of heavy-element masses are based on the fit provided by \citet{2016ApJ...831...64T}, who suggest that a Jupiter-mass planet, on average, has 58 Earth masses of heavy elements. However, the free-floating planet population may originate from the outer disc regions, where the heavy-element content of the planets is lower by a factor of a few. In this case, there would be no issue with the overall mass budget (see also \citealt{2025A&A...693A.302S}).

Within our solar system, Jupiter's heavy-element content is derived through complex interior models (e.g. \citealt{2017GeoRL..44.4649W, 2018A&A...610L..14V, 2022Icar..37814937H}). Current estimates place Jupiter's core in the range of up to 10 Earth masses, while its total heavy-element content is around 30-50 Earth masses. Within our model this implies that Jupiter formed exterior to the water ice evaporation front. Consequently, even if Jupiter migrated several AU \citep{2011Natur.475..206W}, it did not cross the water ice line in this process.

Our simulations show that small ($R_0=50$ AU) discs around solar mass stars can form objects in the brown dwarf mass regime, defined here as 13 Jupiter masses (e.g. Fig~\ref{fig:tracks}). These objects then harbour super-stellar atmospheric abundances (Figs.~\ref{fig:MZplanet} and~\ref{fig:OCNplanet}). In contrast, observations seem to indicate that close-in brown dwarfs may have metallicities corresponding to those of their host star (e.g. \citealt{2024ApJ...977L..32X}). We note that our model can also produce brown dwarfs with sub-stellar metallicities, but such objects would have to be brought closer to their host star via scattering mechanisms. We therefore suggest that the bulk metallicity, in combination with the atmospheric abundances, could indicate whether close-in brown dwarfs formed through the same channels as giant planets or if they exhibit a preference for scattering events.

\section{Summary and conclusion}
\label{sec:conclusion}

In this work we studied the origin of the heavy-element content of giant planets. We used the \textsc{chemcomp} code \citep{2021A&A...654A..71S}, which includes pebble drift and evaporation as well as planetary growth via pebble and gas accretion. Due to the velocity difference between pebbles and gas, the inner disc can be enriched with vapour originating from evaporating pebbles to largely super-stellar values (Fig.~\ref{fig:VaporRadius}). As the planets grow and migrate through the disc, they can accrete this vapour and consequently accrete large amounts of heavy elements (see also \citealt{2021A&A...654A..71S, 2026PASJ...78..493O}). In our study, we did not take the formation and accretion of planetesimals into account, as their formation and accretion would result in a lower total heavy-element content of giant planets \citep{2023A&A...679L...7D, 2026ApJ..1007...19B}. However, here, we are interested in the maximal heavy-element content the planets could reach. Along the same line of thought, we modelled only the growth and evolution of single planets in discs, which therefore have access to the full heavy-element content of the disc (although multiple sequentially forming planets only minimally influence the heavy-element content of the innermost planet, \citet{2024A&A...691A..50E}).

\begin{figure}
 \centering
 \includegraphics[scale=0.6]{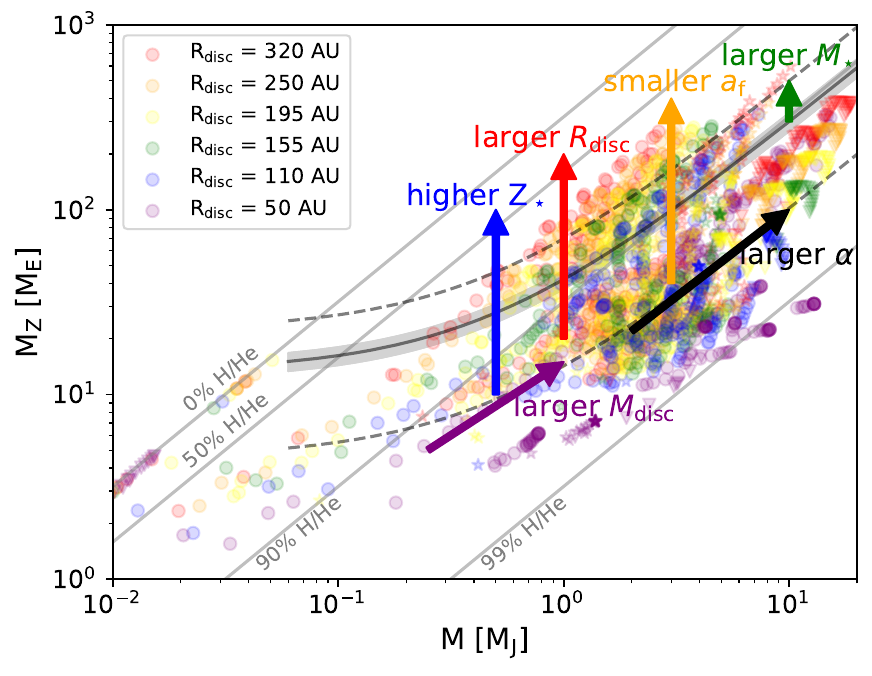}
 \caption{Cartoon illustrating the general trends (arrows) from our simulations for the correlation between total heavy-element mass and total planetary mass. As the trend with stellar mass is very weak, the corresponding arrow is much smaller than the others. The colour-coded symbols are the same as in Fig.~\ref{fig:MZplanet}.
   \label{fig:cartoon}
   }
\end{figure}

In particular we focused on planets forming in discs with different sizes, masses, and viscosities around stars of different stellar masses. Discs with larger sizes can retain a higher heavy-element content in their inner regions for longer (Fig.~\ref{fig:VaporRadius}), while the stellar mass only plays a minor role (Fig.~\ref{fig:Vapor}). This has implications for the heavy-element content of the gas-accreting planets in these discs. Our main findings are illustrated in Fig.~\ref{fig:cartoon} and summarized below:

\begin{itemize}

 \item Planets forming in the inner regions (interior to the CO$_2$ and H$_2$O evaporation fronts) around stars with smaller discs contain fewer heavy elements compared to planets (of the same total mass) forming around stars with larger discs (Fig.~\ref{fig:MZplanet}). This is because inner discs can retain their vapour enrichment for longer if the discs are larger. This occurs because the pebble flux can be maintained for longer in larger discs.

 \item We find that planets forming in discs with the same total mass and size but around stars of higher masses harbour slightly higher heavy-element content compared to planets forming around lower-mass stars (Fig.~\ref{fig:tracksDisc}). This is because discs around higher-mass stars can retain their heavy-element content in the vapour phase for slightly longer (Figs.~\ref{fig:Vapor} and~\ref{fig:VaporRadius}). Together, we find a stronger dependency of the planetary heavy-element content on disc size than stellar mass.

 \item Generally, we find that planets harbouring a heavy-element mass above the observationally derived heavy-element masses (e.g \citealt{2016ApJ...831...64T, 2025ApJ...994...43C, 2026A&A...706A..50P}), must have migrated interior of the water evaporation front (Fig.~\ref{fig:MZdistance}). However, planets with a lower heavy-element mass (or content) could have migrated inwards as well, particularly around stars with smaller disc size (Fig.~\ref{fig:MZplanet}) or in high-viscosity environments. Planets accreting material only exterior to the water ice line, on the other hand, always harbour heavy-element content below the observationally derived mean (Fig.~\ref{fig:MZdistance}). This trend becomes particularly clear for the O/H and C/H in the atmosphere (Fig.~\ref{fig:OCNplanet}).

 \item Metal-rich stars can form more metal-rich planets, but the overall heavy-element content is determined predominantly where the planets accrete material within the protoplanetary disc. If a planet around a metal-rich star accretes material only exterior to the water ice line, it remains less enriched than a giant planet forming around a metal-poor star, provided that the planet accretes most of its material interior to the water ice line (Fig.~\ref{fig:tracksZ}). The trend reported in \citet{2019AJ....158..239T}, namely that more metal-rich stars do not necessarily form more metal-rich planets, thus implies a mixture of the origin of giant planets. They either form in or migrate to the inner disc, or they form in the outer disc and are then scattered inwards after the gas-disc phase, implying different potential obliquities for metal-rich versus metal-poor planets.

\end{itemize}

Our simulations show a complex interplay between stellar mass, metallicity, disc size, and formation location to explain the high heavy-element content of giant planets (Fig.~\ref{fig:cartoon}). Current observations are limited by small sample sizes making it difficult to draw a definite conclusion at the moment. We expect that a combination of stellar metallicity determinations via 4MOST \citep{2012SPIE.8446E..0TD} and planetary detections via PLATO \citep{2014ExA....38..249R} will help verify our predictions about the origin of the higher heavy-elements mass of giant planets in the coming years.

\begin{acknowledgements}

B.B., thanks Ravit Helled for thoughtful discussions and feedback on the manuscript. B.B. thanks the anonymous referee for their insightful comments that helped to improve the quality of the manuscript.

\end{acknowledgements}

\bibliographystyle{aa}
\bibliography{Stellar}

\newpage

\begin{appendix}

\clearpage

\section{Disc compositions}
\label{ap:discs}

\begin{figure*}
 \centering
 \includegraphics[scale=0.375]{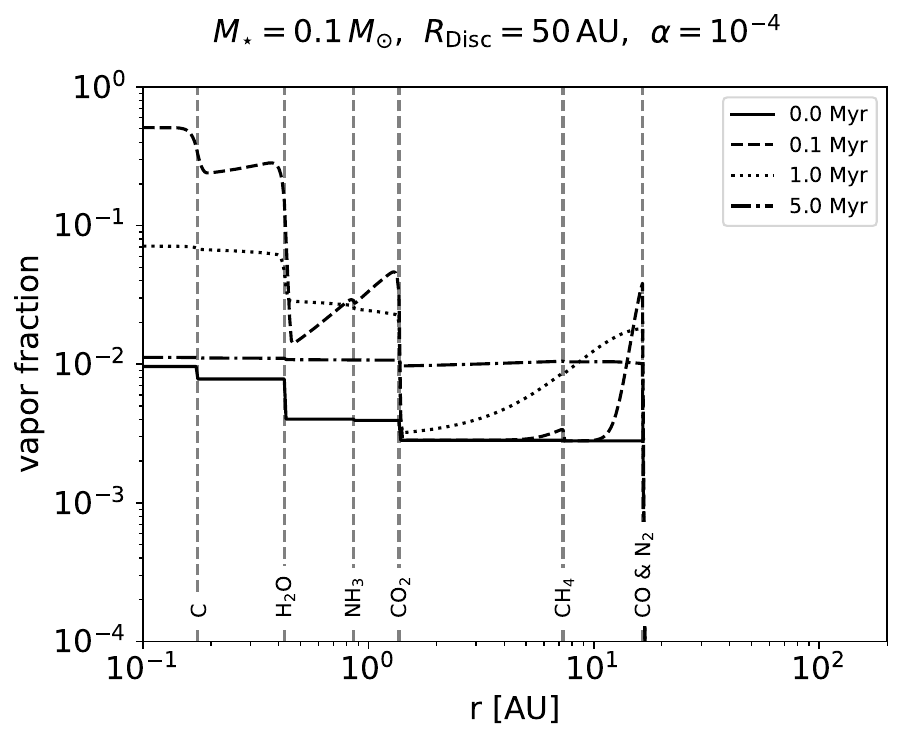} \quad
 \includegraphics[scale=0.375]{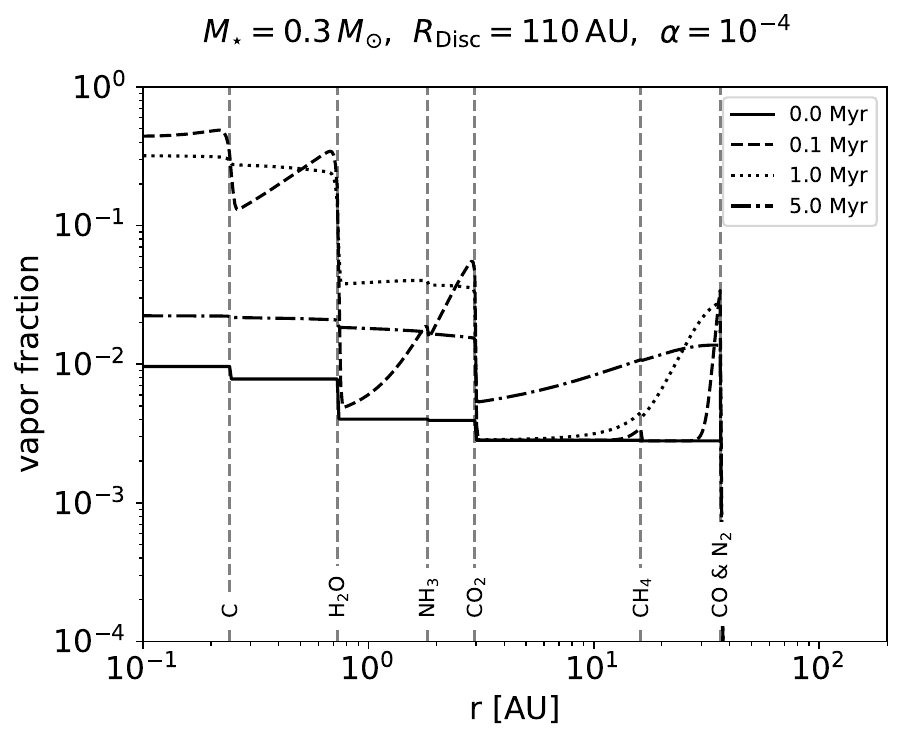} \quad
 \includegraphics[scale=0.375]{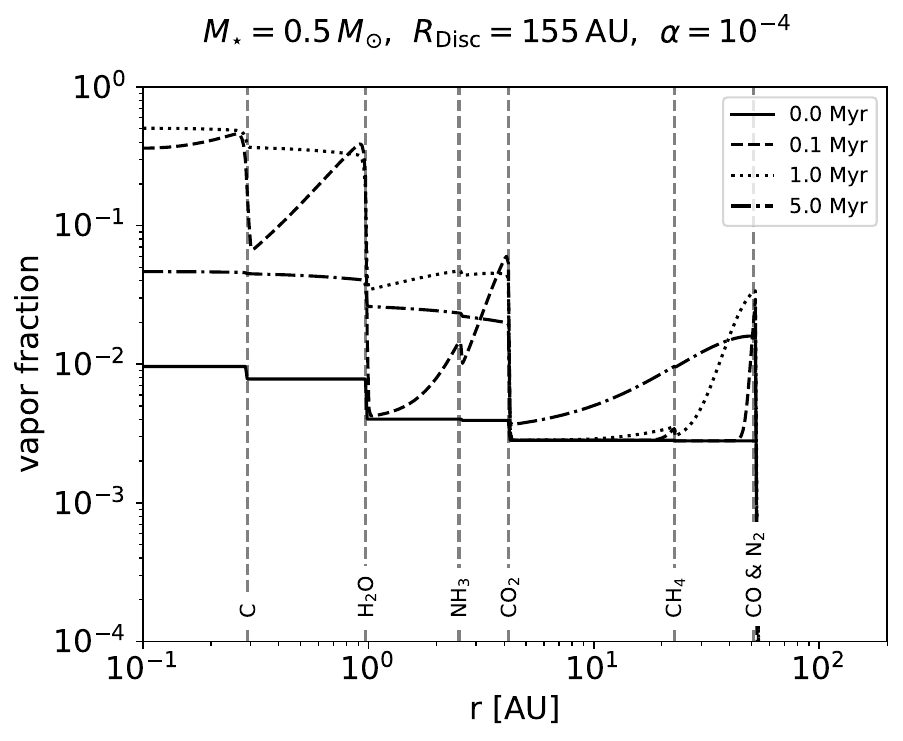} \quad
 \includegraphics[scale=0.375]{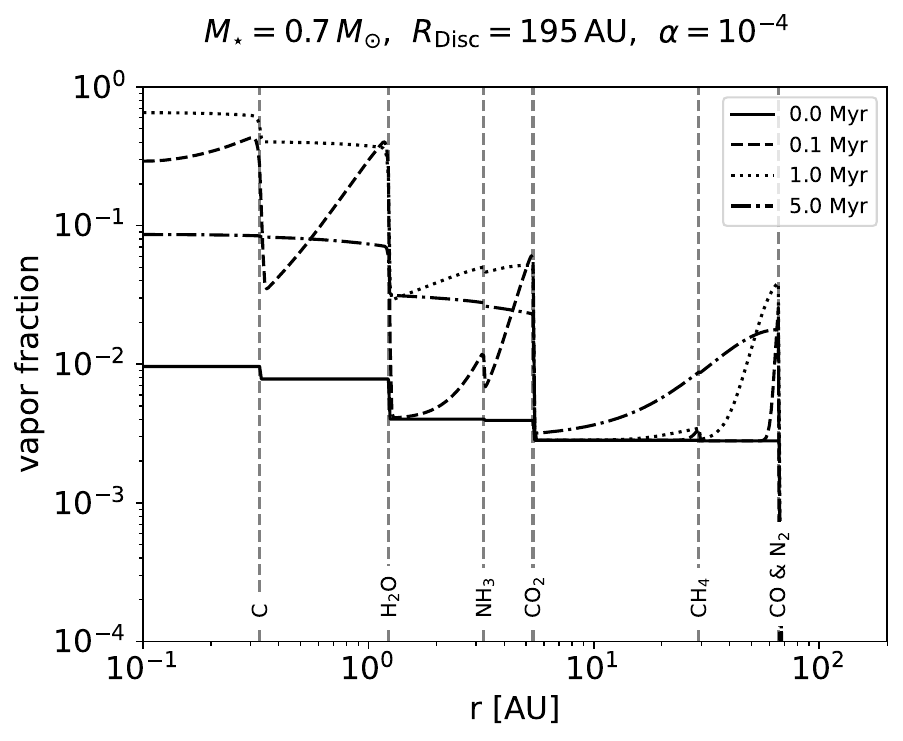} \quad
 \includegraphics[scale=0.375]{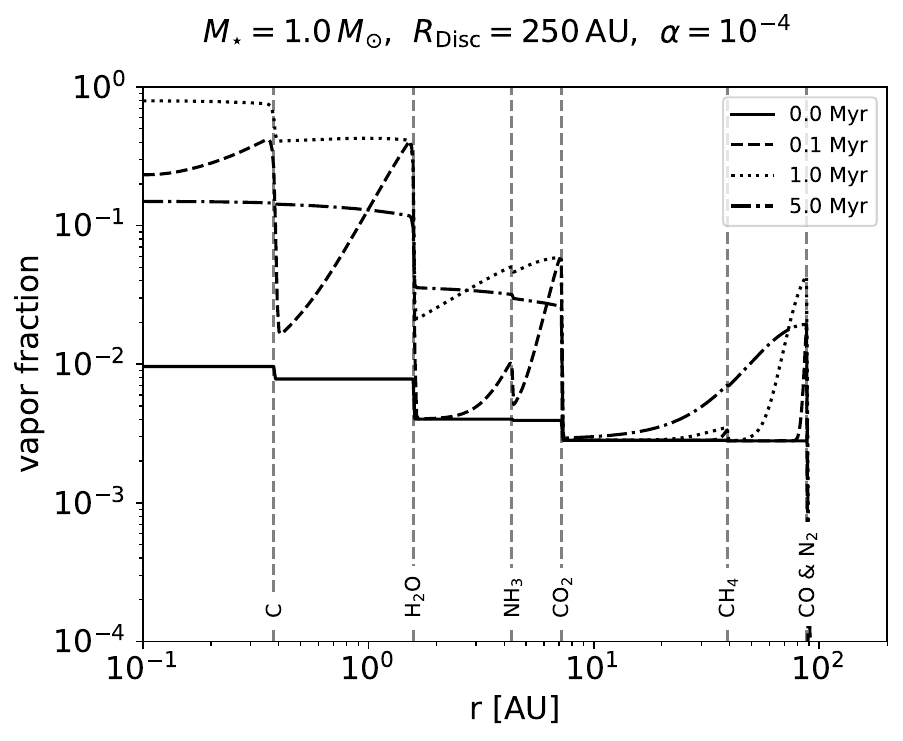} \quad
 \includegraphics[scale=0.375]{VaporM14_320AU.pdf}
 \caption{Vapour fraction within protoplanetary discs with $\alpha=10^{-4}$ as a function of time for the different stellar masses, using a scaling of the disc size with stellar mass. The disc sizes correspond to the same disc sizes as in Fig.~\ref{fig:VaporRadius}.
   \label{fig:Vapor}
   }
\end{figure*}

We show in Fig.~\ref{fig:Vapor} the vapour fraction within discs around stars with different masses. The disc sizes are assumed to scale with the stellar mass as in \citet{2023arXiv230815128M} to illustrate the effect of different stellar masses on the disc evolution and correspond to the same sizes as in Fig.~\ref{fig:VaporRadius}. As smaller stars are less luminous than higher-mass stars, their evaporation fronts are located closer to the host star. This results in a slightly faster depletion of the vapour content in the inner discs around lower-mass stars. We emphasize that the faster decrease in vapour content is mainly driven by the smaller disc size (see Fig~\ref{fig:VaporRadius}), while the effect of stellar mass is much weaker. For the smallest disc size, we find a depletion that is around a factor of 2-3 faster around the 0.1 solar-mass star compared to the 1.4 solar-mass star. This shows that the vapour content in the disc primarily depends on the disc size and the effect of the stellar mass is only a second-order effect.

\begin{figure*}
 \centering
 \includegraphics[scale=0.375]{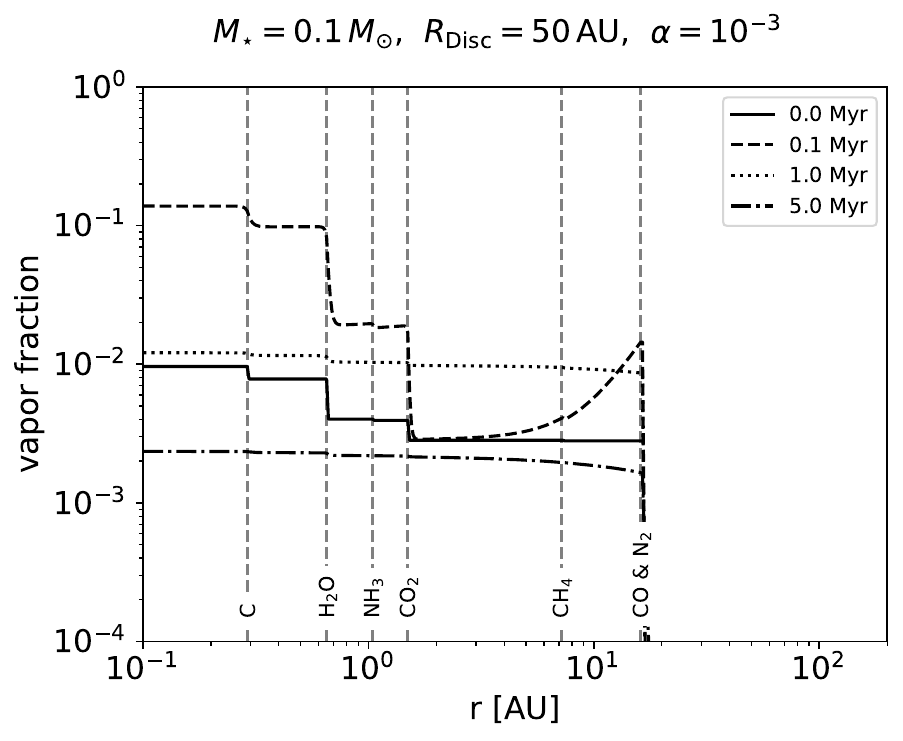} \quad
 \includegraphics[scale=0.375]{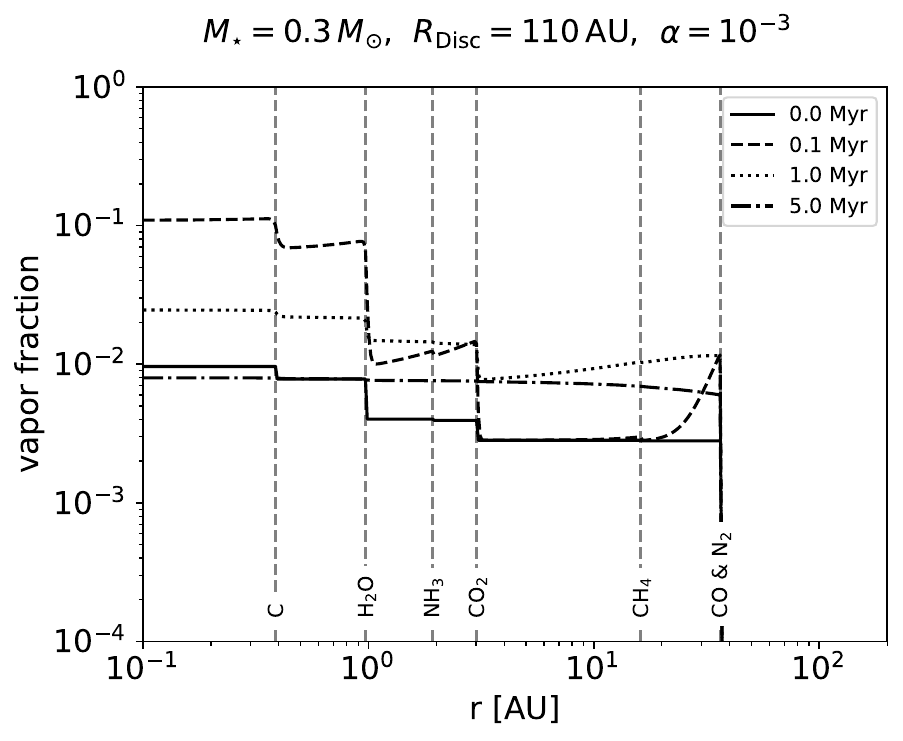} \quad
 \includegraphics[scale=0.375]{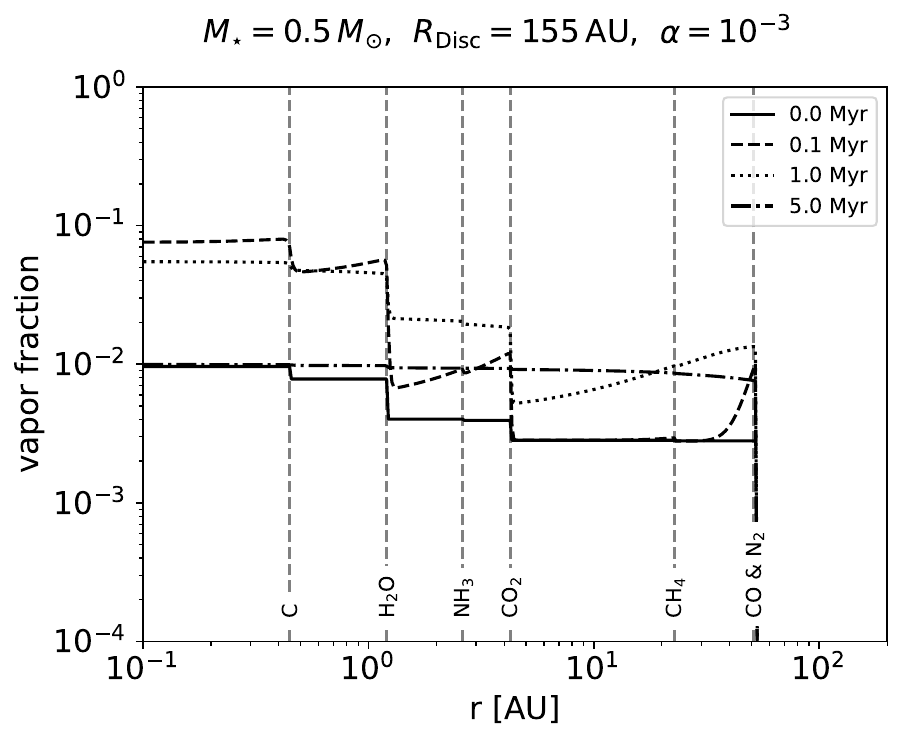} \quad
 \includegraphics[scale=0.375]{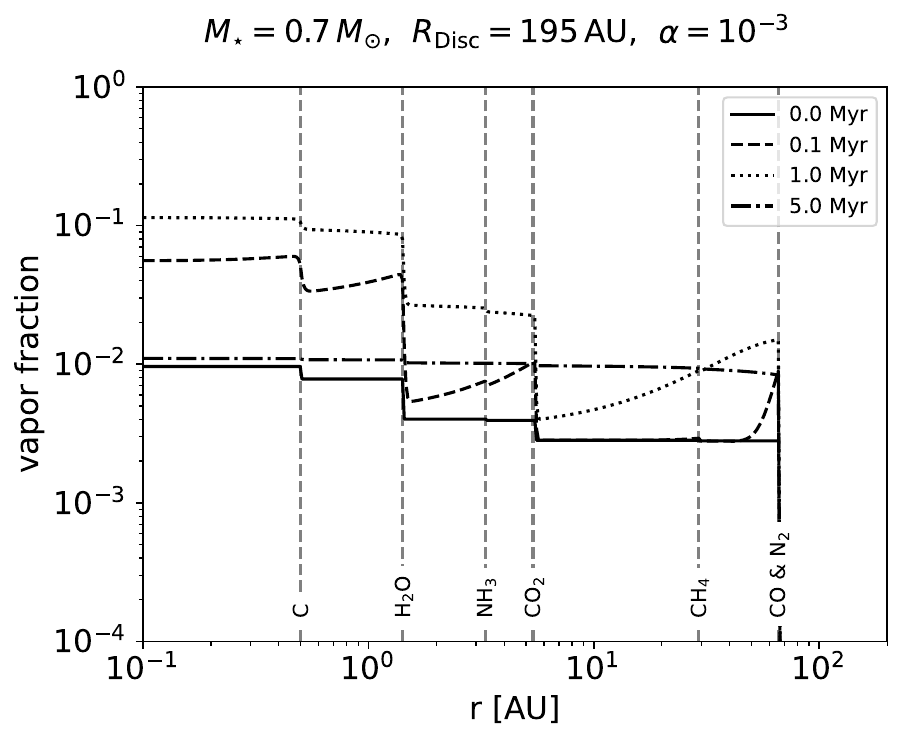} \quad
 \includegraphics[scale=0.375]{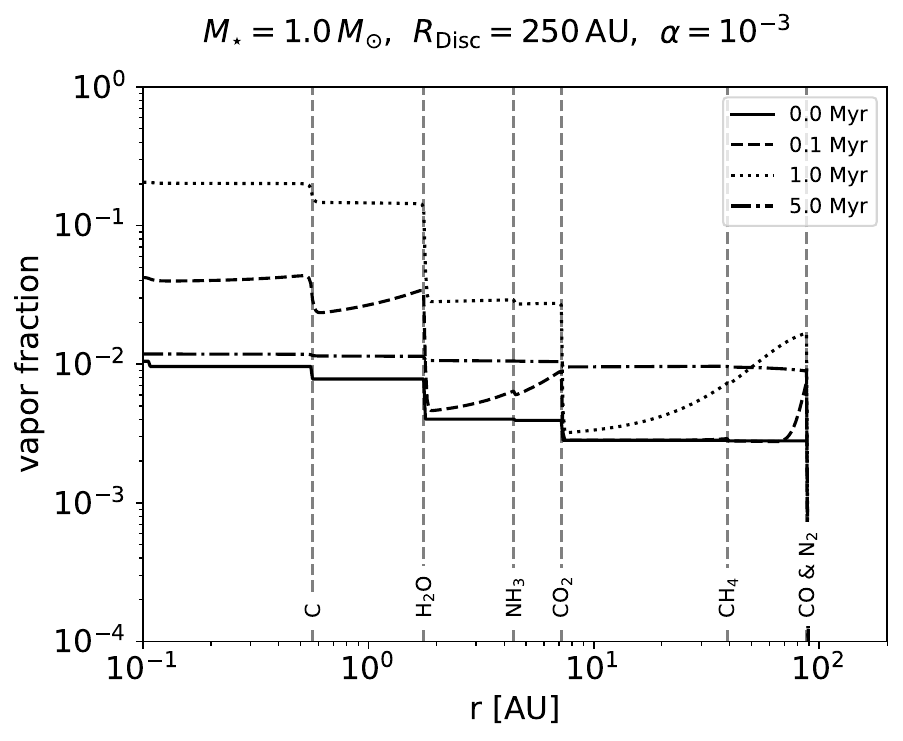} \quad
 \includegraphics[scale=0.375]{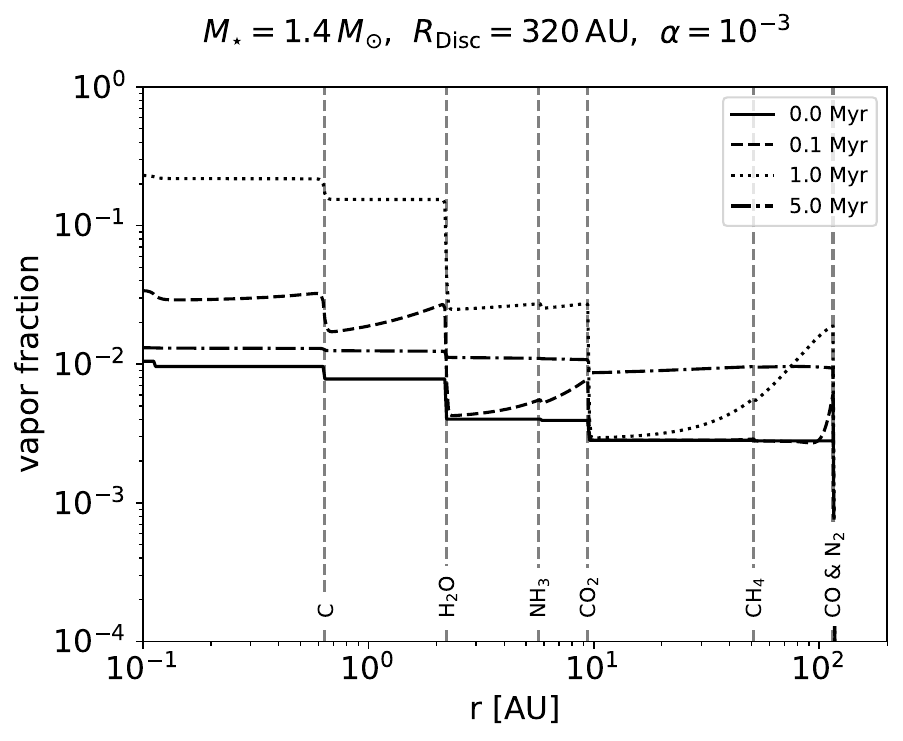}
 \caption{Vapour fraction within protoplanetary discs with $\alpha=10^{-3}$ as a function of time for the different stellar parameters. The vapour fraction is defined as the mass fraction of the sum of C, O, and N compared to the background gas consisting of hydrogen and helium.
   \label{fig:Vapor1e-3}
   }
\end{figure*}

We show in Fig.~\ref{fig:Vapor1e-3} the time evolution of the vapour fraction of discs around stars with different stellar masses and disc sizes with $\alpha=10^{-3}$. The higher disc viscosity results in a faster decay of the vapour content in the inner disc due to two effects: (i) the higher viscosity limits the size of the pebbles in the fragmentation regime (e.g. \citealt{2012A&A...539A.148B}), reducing radial drift and, consequently, the initial enrichment; and (ii) the higher viscosity enables faster accretion of vapour onto the star, depleting the vapour content in the disc. As seen in discs with lower viscosities (Fig.~\ref{fig:Vapor}), this effect is more efficient for lower-mass stars with small disc sizes, where vapour enrichment can even drop below the initial vapour fraction by the end of the disc's lifetime. Nevertheless, the general trend -- whereby inner discs around stars with larger disc sizes remain enriched for a longer time -- also holds at higher viscosities.

\section{Influence of outer gaps}
\label{ap:gaps}

Outer gaps in protoplanetary discs can block inward-flowing pebbles (e.g. \citealt{2006A&A...453.1129P, 2012A&A...538A.114P}), which will result in lower vapour enrichment in the inner disc regions (e.g. \citealt{2023arXiv230701789K, 2024A&A...686L..17M}). The blocking ability of the gaps depends on their depths as well as when during the disc evolution they appear. We include the gap into the protoplanetary disc via our previous model \citep{2024A&A...686L..17M}.

We introduce the gap profile at a specific time $t_{\rm gap}$ by modifying the $\alpha$-viscosity parameter in the disc to include a numerical factor that describes the gap profile.
\begin{equation}
 \mathcal{N}(r) = 1 - [1-f_{\rm gap}] \exp \left( \frac{1}{2} \left[ \frac{r-r_{\rm gap}}{\sigma} \right]^2 \right) \ ,
\end{equation}
where $0 \le f_{\rm gap} \le 1$ is the gap depth (lower values mean deeper gaps), $r_{\rm gap}$ is the location of the gap and $\sigma$ is the width of the gap given as
\begin{equation}
 \sigma = 2H_{\rm g} (r=r_{\rm g}) / [2 \sqrt{2 \log (2)}] \ .
\end{equation}
$H_{\rm g} (r=r_{\rm g})$ corresponds to the gas scale height at the position of the gap. The resulting $\alpha$ disc profile is then
\begin{equation}
 \alpha' = \alpha / \mathcal{N}(r) \ ,
\end{equation}
where we only apply the modifications via $\alpha$ on the gas surface density. The original value of $\alpha$ is used for the dust processes.

\begin{figure}
 \centering
 \includegraphics[scale=0.6]{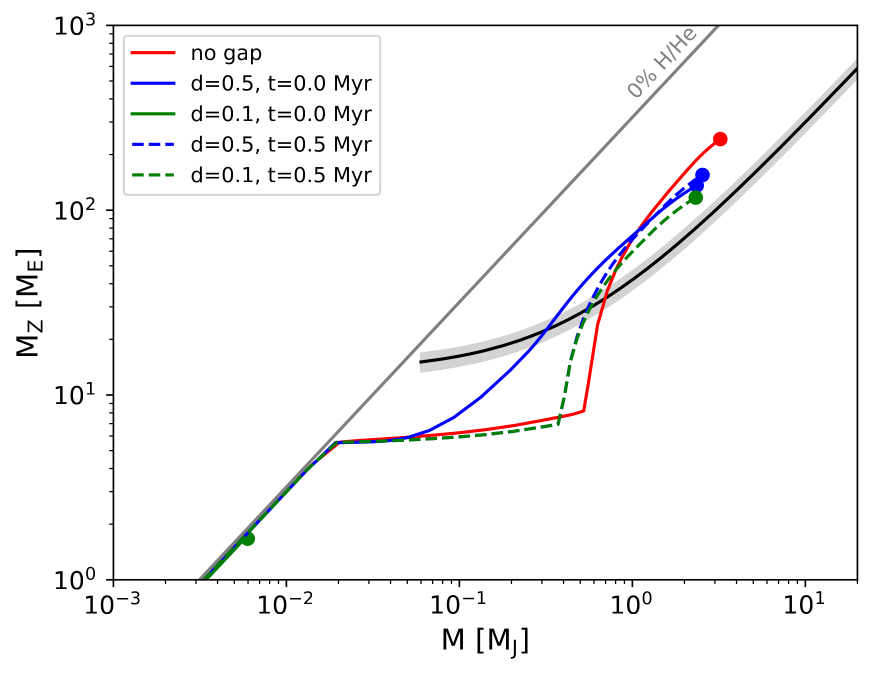}
 \caption{Heavy-element mass as a function of planetary mass for planets forming around a 1.4 $M_\odot$ star with a disk size of 320 AU, $\alpha=10^{-4}$, and gaps introduced at different times with varying strength. The gaps are always at 50 AU. The planets start at 2 AU.
   \label{fig:MZgaps}
   }
\end{figure}

We show in Fig.~\ref{fig:MZgaps} the evolution of the heavy-element mass of planets forming in discs containing outer gaps at 50 AU. We introduced gaps with depths of 0.5 and 0.1 either directly at the beginning of the simulation or after 0.5 Myr. Introducing the gap at the beginning of the simulations has the biggest effect, because the planet forming in a disc with a gap depth of 0.1 does not grow very large. This is caused by the fact that the deep gap blocks inward-drifting pebbles efficiently and the planet thus does not grow. A slightly shallower gap depth allows pebbles to still drift inwards and thus growth of the planet. However, the planet grows slower compared to the nominal case and it thus migrates inwards more efficiently before it transitions into the gas accretion phase. Consequently, it crosses the water ice line a bit earlier and starts to accrete water vapour relatively early during its formation and thus harbours at its early stages a slightly higher heavy-element content compared to the planet forming in a disc without a gap. However, in the end, the planetary heavy-element content is lower compared to that of a planet forming in a disc without a gap, because the pebble flux into the inner disc is reduced, likewise reducing the vapour content in the disc.

In contrast, if the gap forms at a later time ($t_{\rm gap}=0.5$ Myr), there is not significant effect on the growth of the planetary core, because the pebbles can flow freely for the first 0.5 Myr. However, over the course of 5 Myr, some pebbles are blocked, and the heavy-element content remains lower compared to the planet forming in the simulation without gaps. This shows that outer gaps can indeed reduce the maximal heavy-element content of planets or even hinders giant planet formation all together, especially if the gap forms early\footnote{Gaps located closer in further reduce the pebble flux, resulting in less efficient giant planet growth and lower enrichment.}. Thus, in order to achieve the maximum heavy-element content within the pebble drift and evaporation scenario, discs without (initial) gaps are preferred.

\section{Disc lifetime}
\label{ap:lifetime}

\begin{figure}
 \centering
 \includegraphics[scale=0.6]{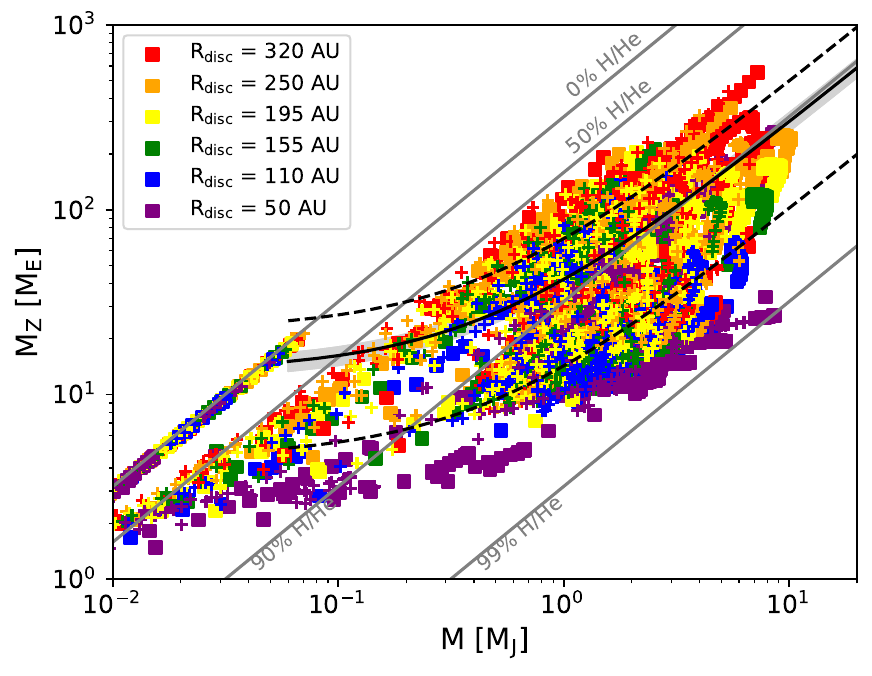}
 \caption{Heavy-element mass as a function of planetary mass for our simulations. We show here the data after 2 Myr (plus markers) and 3 Myr (square markers) to illustrate the evolution of the heavy-element content of the planets as a function of time.
   \label{fig:MZyoung}
   }
\end{figure}

We show in Fig.~\ref{fig:MZyoung} the heavy-element mass of planets in discs whose lifetime is only 2 or 3 Myr. We note that we use the same simulations as in the main paper, and instead plot the resulting heavy-element masses of the planets after 2 and 3 Myr.

A reduced disc lifetime results in less time for the planet to accrete gas, giving rise to lower planetary masses (on average). This effect allows us to fill the parameter space of planets below 0.7 Jupiter masses, where our nominal simulations with a disc lifetime of 5 Myr were not able to produce many planets. This shows that our model can be applied to the whole mass range of planets. We note that our simulations allow us to reproduce the shape of the derived heavy-element masses as a function of planetary mass of \citet{2025ApJ...994...43C}.

Our simulations also show an empty region between 0\% H/He and 50\% H/He fraction. This is caused by the transition into runaway accretion. Once the planet reaches its pebble isolation mass, it accretes its envelope relatively fast, so that its mass doubles, but it does not necessarily increase its heavy-element content, which, at this stage, remains dominated by the planetary core.

\section{Atmospheric abundances}
\label{ap:abu}

\begin{figure*}
 \centering
 \includegraphics[scale=0.37]{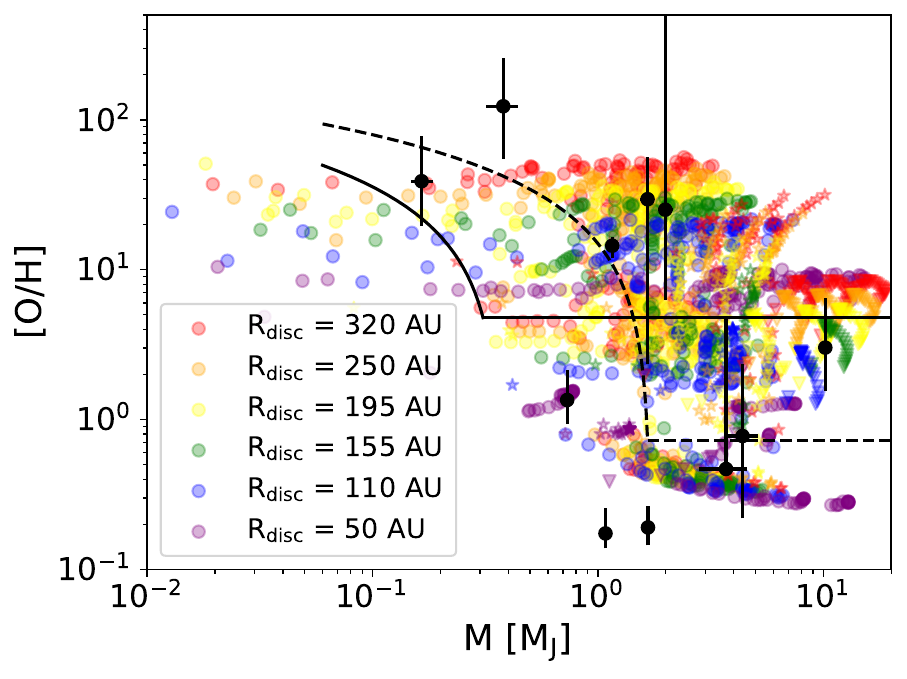} \quad
 \includegraphics[scale=0.37]{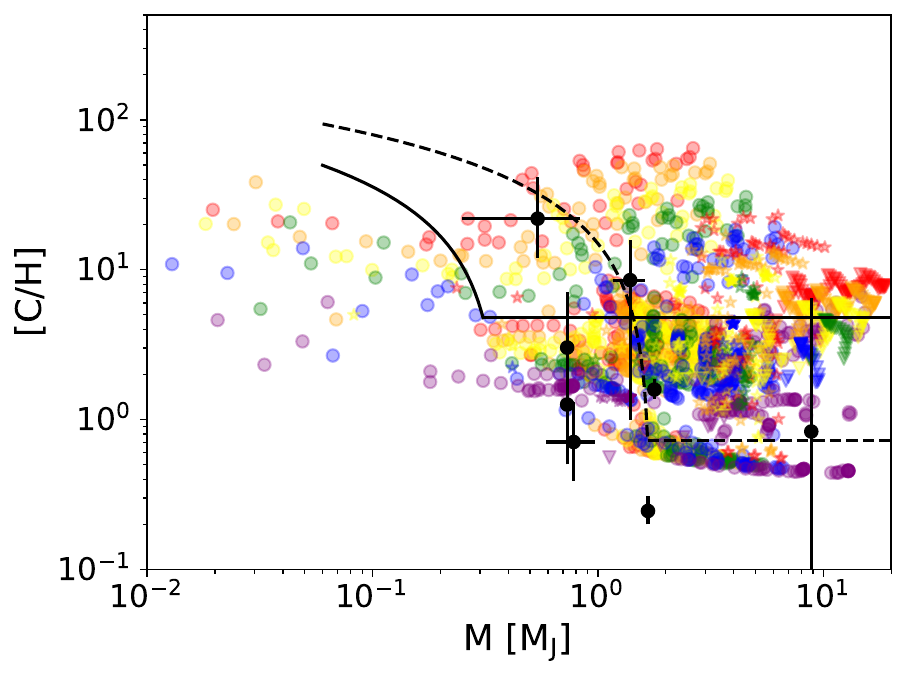} \quad
 \includegraphics[scale=0.37]{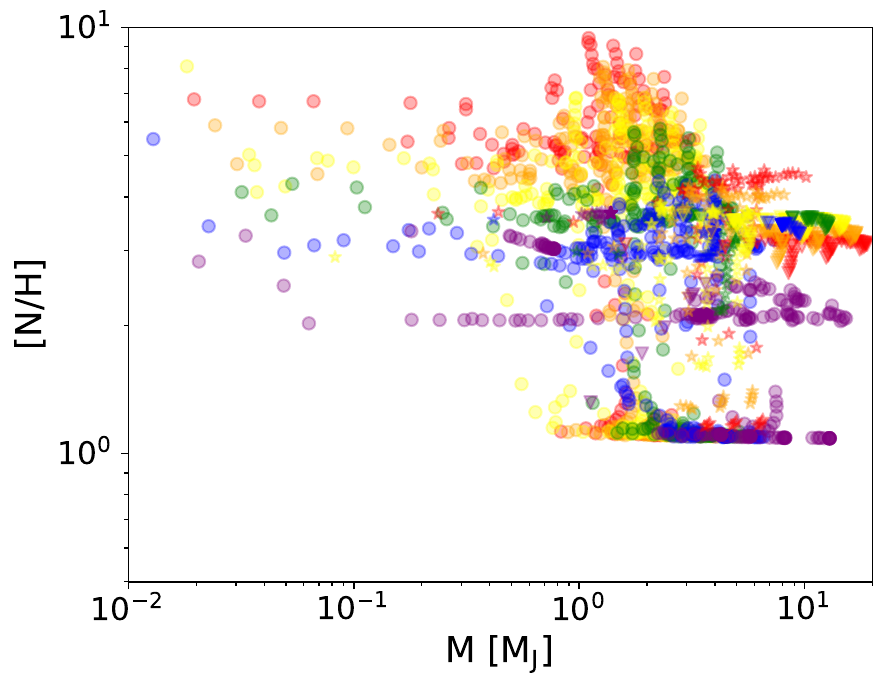} \\ \hspace{0.3cm}
 \includegraphics[scale=0.4]{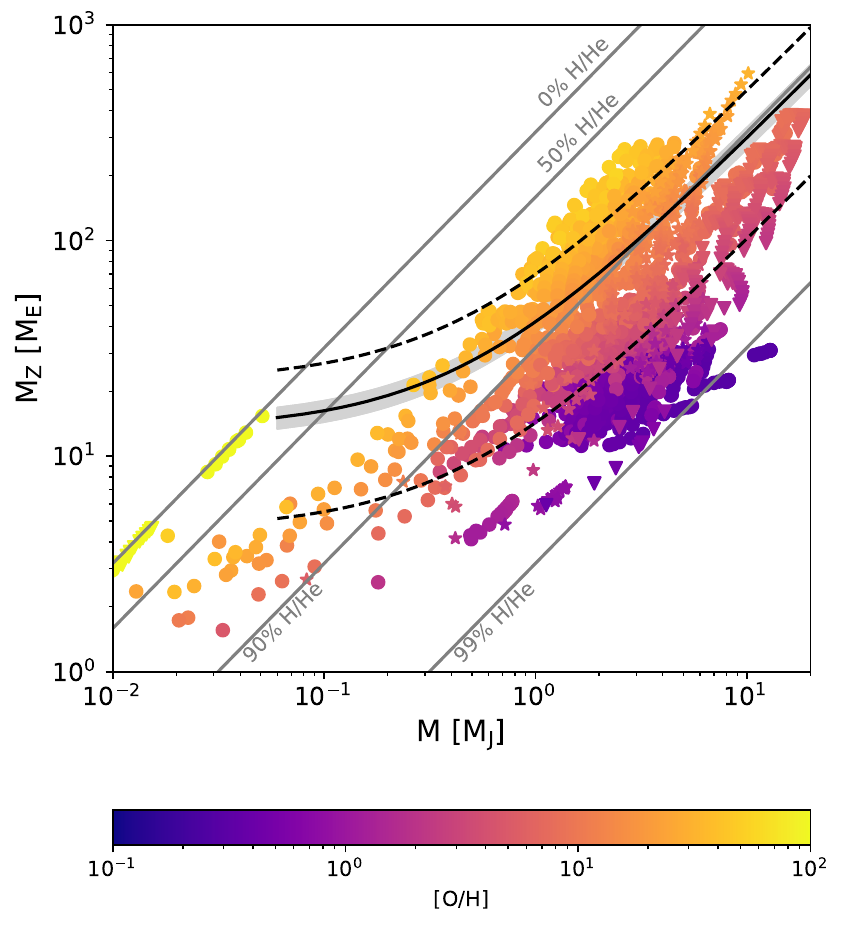} \quad
 \includegraphics[scale=0.4]{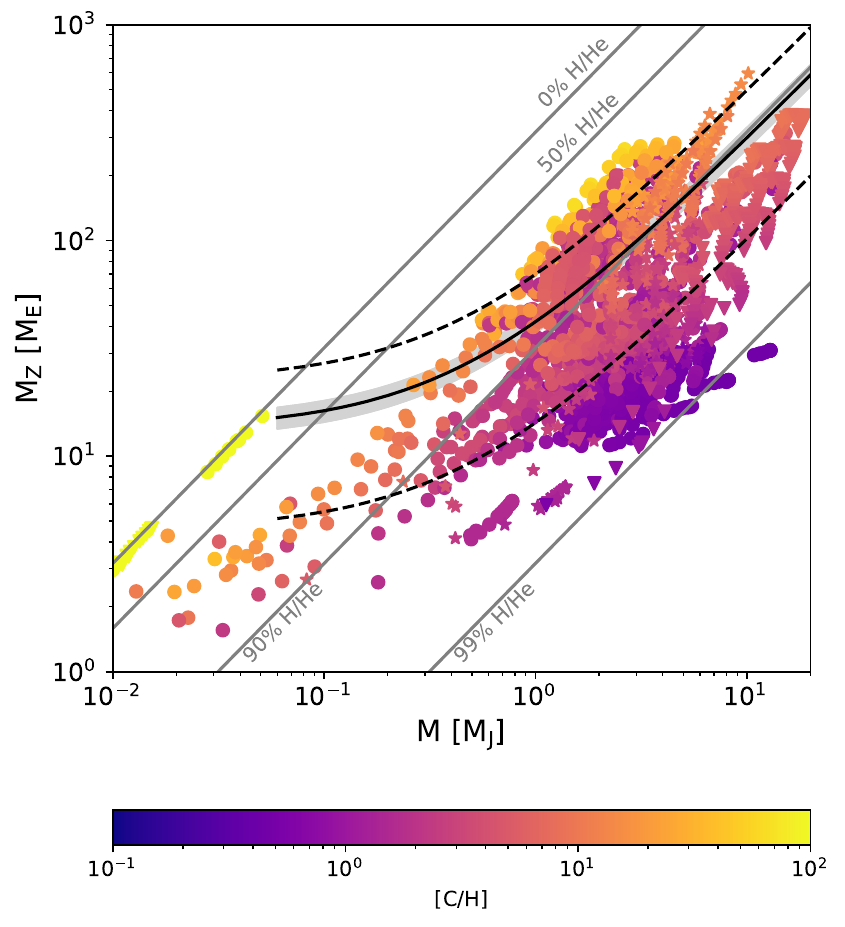} \quad
 \includegraphics[scale=0.4]{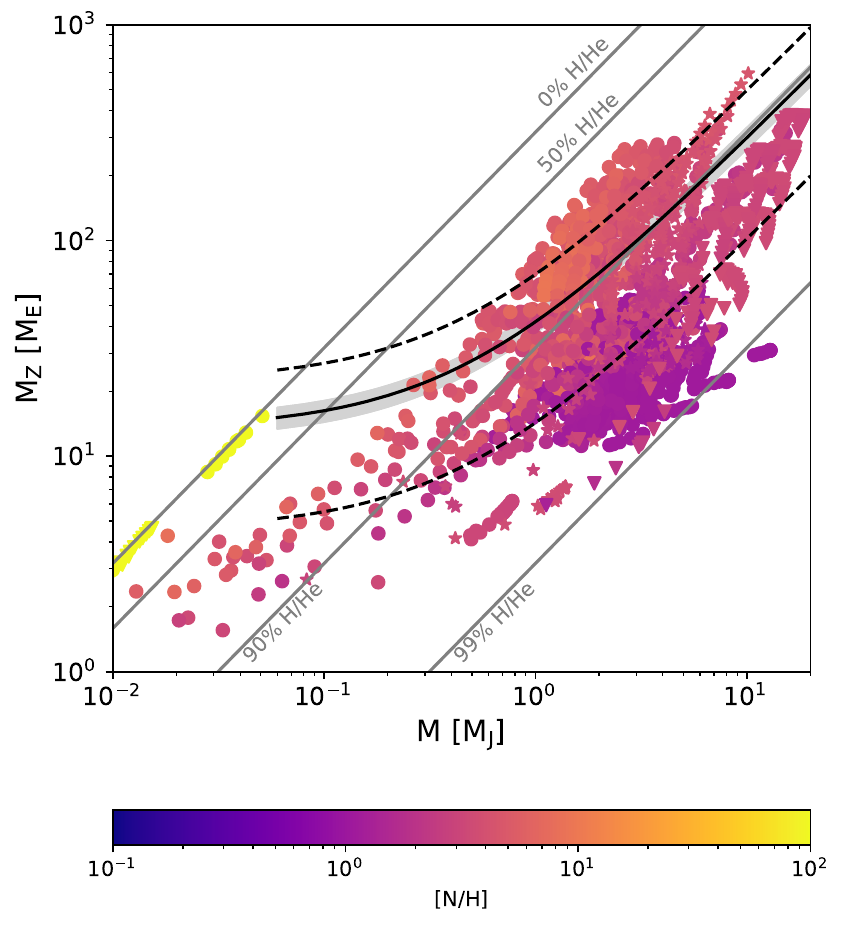}
 \caption{Top: O/H, C/H, and N/H of the atmospheres of all simulated planets for with $Z=0.016$. The circles correspond to $\alpha=10^{-4}$; the stars correspond to $\alpha=3 \times 10^{-4}$; and the upside down triangles correspond to $\alpha=10^{-3}$. The lines and data points correspond to the fits and observational data from \citet{2026AJ....171...31L}, where the solid line corresponds to the fits of the data from transmission spectroscopy and the dashed line corresponds to eclipse spectroscopy. Note the different scaling on the y-axis for the plots. Bottom: Heavy-element mass of planets as a function of their total mass, colour-coded by the atmospheric abundances of O/H, C/H, and N/H.
   \label{fig:OCNplanet}
   }
\end{figure*}

We show in Fig.~\ref{fig:OCNplanet} the atmospheric O/H, C/H and N/H ratios for all planets formed in our simulations. The high heavy-element content of the giant planets in our simulations originates from the accretion of vapour-enriched gas, visible via the super-stellar abundance measurements of the volatiles. In particular the O/H and C/H ratio can reach values larger than 10 times stellar, resulting in high heavy-element masses of the planets. Furthermore, the maximum enrichment that the planets can reach in O/H and C/H depends on the disc's viscosity. Planets forming in discs with lower viscosity can reach a higher maximal C/H and O/H value compared to planets forming in higher viscosity discs. For example, planets forming in discs with $\alpha=10^{-3}$ are, at most, only enriched to a few times the stellar oxygen and carbon abundances, while planets forming in discs with $\alpha=10^{-4}$ can be enriched up to a few tens of times the stellar values. These extreme values are in line with observationally derived atmospheric abundances of giant exoplanets (e.g. \citealt{2025A&A...699A.342B, 2025NatAs...9..845E, 2026AJ....171...31L}).

As discussed above, planets forming around stars with larger disc sizes can have higher heavy-element masses. This trend is reflected in the planetary O/H and C/H ratios: planets forming around stars with larger disc sizes can have much greater C/H and O/H ratios compared to their counterparts forming around stars with smaller disc sizes. This difference -- as discussed above -- is due to the more efficient accretion of vapour-enriched gas in discs around stars with larger sizes.

Regardless of disc size, we find that low-mass planets (above 0.7 Jupiter masses) can harbour greater atmospheric O/H and C/H compared to high-mass planets. This is because the lower-mass planets in our simulations form in low-viscosity discs, where the vapour content of the inner disc can remain high during the whole accretion phase of the planets (Fig.~\ref{fig:Vapor}). In contrast, higher-mass planets (above a few Jupiter masses) form in discs with higher viscosities, leading to a depletion in the vapour content of the disc towards the end of the disc's lifetime (Fig.~\ref{fig:Vapor1e-3}), resulting in lower overall enrichment of these more massive planets. This reflects the observational trends discussed in \citet{2026AJ....171...31L}.

We note that our simulations show that planets can be sub-stellar in O/H and C/H, if they form in the outer disc regions and do not accrete large amounts of vapour. However, these planets, even though they are sub-stellar in their atmospheric abundances have a super-stellar metallicity (Fig.~\ref{fig:MZplanet}), due to the contribution of the planetary core to the overall metallicity. We caution though to make a direct comparison between our simulations and the derived observational data, because each planet orbits a star with different composition that influences the composition of the growing planets around it (e.g. \citealt{2022A&A...665A.138B, 2026A&A...706A..30O}; Fig.~\ref{fig:tracksZ}).

While our simulations show a clear trend with O/H and C/H, there is no clearly visible trend with N/H. This is caused by the fact that nitrogen-rich ice mostly evaporates in the very outer disc regions, where it then needs to be transported inwards on a (very long) viscous timescale, erasing the differences between the different disc sizes.

\section{Scaling with disc radius}
\label{ap:discradius}

\begin{figure*}
 \centering
 \includegraphics[scale=0.6]{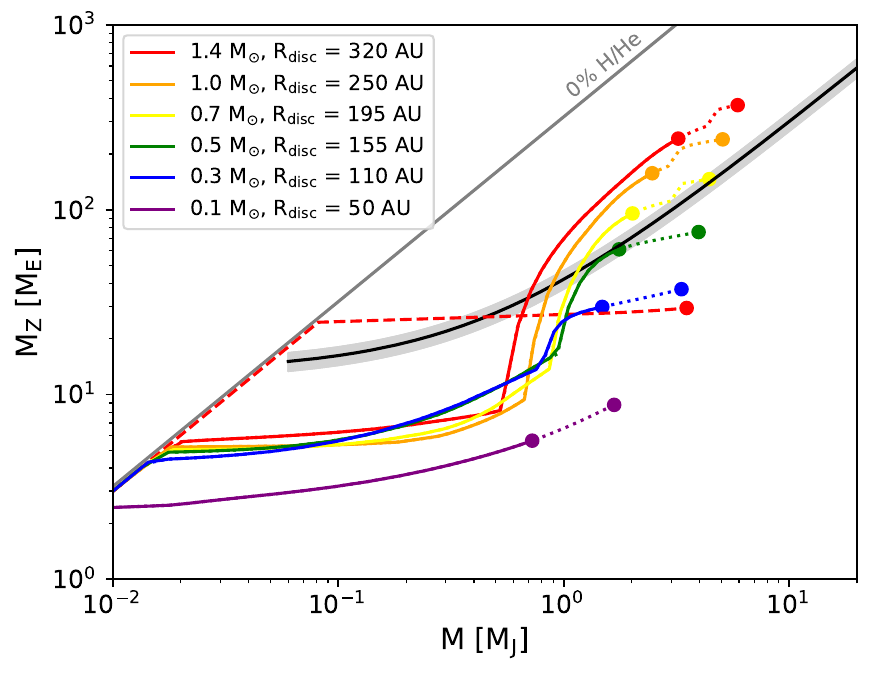} \quad
 \includegraphics[scale=0.6]{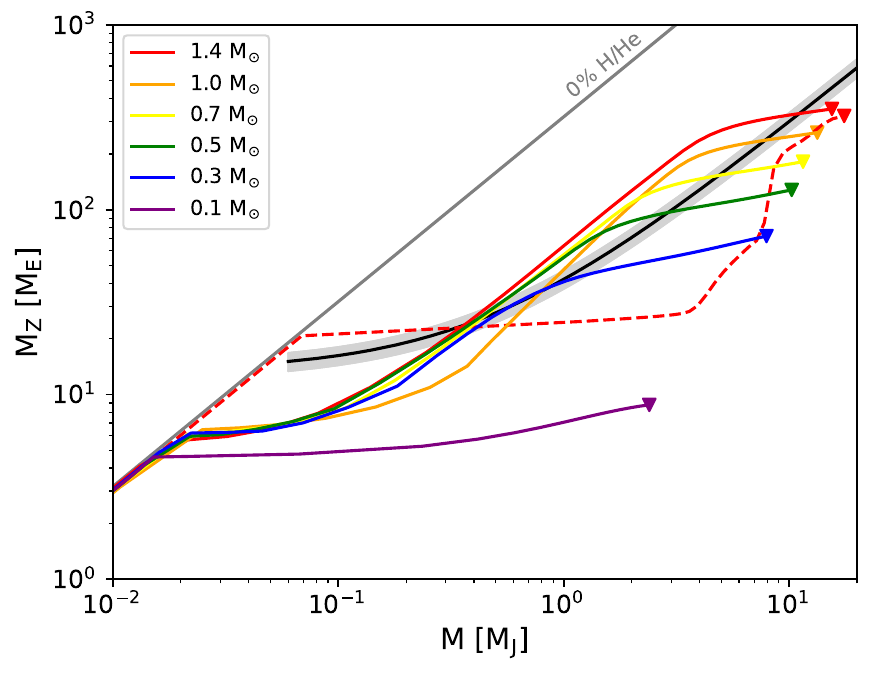}
 \caption{Heavy-element mass evolution of planets starting at 2.0 AU in discs with $\alpha=10^{-4}$ (left) and $\alpha=1 \times 10^{-3}$ (right) around stars with different masses and disc radii up to an evolution of 5 Myr. The dotted lines in the left-hand plot correspond to planets that are evolved for 10 Myr in the disc. The dashed lines correspond to planets starting in systems with 1.4 M$_\odot$ stars in the outer disc at 20 AU, where the pebble isolation mass is higher but vapour enrichment is lower (Fig.~\ref{fig:Vapor}).
   \label{fig:tracksstar}
   }
\end{figure*}

We show in Fig.~\ref{fig:tracksstar} the heavy-element evolution of planets forming around stars with different masses and disc sizes as well as in discs with different viscosities. The disc sizes are assumed to scale with the stellar mass, as in our previous work \citep{2023arXiv230815128M}. The heavy-element mass of the formed planets is - under the assumption that the disc size reduces with stellar mass - now a clear function of stellar mass. This trend would explain the observations of \citet{2025A&A...693L...4M}, where giant planets around lower-mass stars harbour lower heavy-element content. As this trend is not recovered in \citet{2025ApJ...994...43C}, it would imply from our simulation perspective that the disc size is not scaling strongly with the stellar mass, as also indicated by observations \citep{2020ApJ...895..126H, 2022ApJ...931....6L}. However, not all stars have discs of the same size, which can thus explain the diversity of the formed giant planets.

\end{appendix}

\end{document}